\documentclass[12pt]{article}
\pdfoutput=1

\usepackage{amsmath,amssymb,amsfonts,graphicx,amsfonts}
\usepackage{textcomp,mathcomp}
\usepackage{color}
\usepackage[hidelinks]{hyperref}
\usepackage{ascmac}
\usepackage{physics}
\usepackage{dcolumn}
\usepackage{bm}
\usepackage{caption}
\usepackage{subcaption}
\usepackage[utf8]{inputenc}
\usepackage{booktabs}
\usepackage{xcolor}

\allowdisplaybreaks[4]

\date{}
\begin{document}
\begin{flushright}
\today\\
RIKEN-iTHEMS-Report-24
\end{flushright}

\vspace{0.1cm}

\begin{center}
  {\Large Variational Monte Carlo with Neural Network Quantum States}\\
  \vspace{3mm}
  {\Large  for Yang-Mills Matrix Model}\\

\end{center}
\vspace{0.1cm}
\vspace{0.1cm}
\begin{center}
{\large
Norbert Bodendorfer$^\mu$, Onur Oktay$^\sigma$, Vaibhav Gautam$^{\gamma}$,\\
Masanori Hanada$^\gamma$, and Enrico Rinaldi$^\delta$
}
\end{center}
\vspace{0.3cm}

\begin{center}

$^\mu$University of Regensburg, Institute of Theoretical Physics\\ 
Universit\"atsstrasse 31, D-93053 Regensburg, Germany

$^\sigma$ Department of Physics, Karamanoglu Mehmetbey University, 70100, Karaman, Turkey

$^\gamma$School of Mathematical Sciences, Queen Mary University of London\\
Mile End Road, London, E1 4NS, United Kingdom

$^\delta$Quantinuum K.K., Otemachi Financial City Grand Cube 3F\\
1-9-2 Otemachi,Chiyoda-ku, Tokyo, Japan\\
$^\delta$ Interdisciplinary Theoretical and Mathematical Sciences (iTHEMS) Program\\
RIKEN, Wako, Saitama 351-0198, Japan\\
$^\delta$Center for Quantum Computing (RQC), RIKEN, Wako, Saitama 351-0198, Japan
\end{center}

\vspace{1.5cm}

\begin{center}
  {\bf Abstract}
\end{center}

We apply the variational Monte Carlo method based on neural network quantum states, using a neural autoregressive flow architecture as our ansatz, to determine the ground state wave function of the bosonic SU($N$) Yang-Mills-type two-matrix model at strong coupling.
Previous literature hinted at the inaccuracy of such an approach at strong coupling.
In this work, the accuracy of the results is tested using lattice Monte Carlo simulations: we benchmark the expectation value of the energy of the ground state for system sizes $N$ that are beyond brute-force exact diagonalization methods.
We observe that the variational method with neural network states reproduces the right ground state energy when the width of the network employed in this work is sufficiently large.
We confirm that the correct result is obtained for $N=2$ and $3$, while obtaining a precise value for $N=4$ requires more resources than the amount available for this work.

\newpage

\tableofcontents

\section{\label{sec:introduction}Introduction}
Yang-Mills theory and matrix models with gauge group SU($N$), where $N=2,3,4,\cdots$, have been playing prominent roles in theoretical physics.
Applications include the systematic understanding of strong dynamics based on the 't Hooft expansion~\cite{tHooft:1973alw} and the nonperturbative study of quantum gravity in the framework of holographic duality~\cite{Banks:1996vh,Itzhaki:1998dd}.
Numerical techniques are valuable tools for the study of those theories, due to their nonperturbative nature.
Traditionally, Markov Chain Monte Carlo (MCMC) methods have been applied successfully to the class of problems accessible via the Euclidean path integral, such as thermodynamics or the spectrum of low-energy excitations.
See e.g., Ref.~\cite{Lucini:2001ej} and Refs.~\cite{Anagnostopoulos:2007fw,Catterall:2008yz} for SU($N$) Yang-Mills theory and the D0-brane matrix model, respectively.

Despite its immense success in numerical nonperturbative physics, the Euclidean path integral falls short in addressing many important issues.
For example, while we can obtain the information from the canonical ensemble, it is not straightforward to access individual quantum states.
Furthermore, even among the problems accessible via the Euclidean path integral, some important problems cannot be studied via MCMC because of the sign problem.
A particularly well-known example is Quantum Chromo-Dynamics (QCD) -- which is SU(3) Yang-Mills coupled to quarks -- at finite baryon density~\cite{Nagata:2021ugx}. 
Therefore, it is important to develop alternative methods. 

In this paper, we consider the Variational Monte Carlo (VMC) method based on Neural Network Quantum States (NNQS)~\cite{10.1126/science.aag2302}.
The NNQS are quantum states expressed by parameterized neural networks and are among the most expressive ansatz for the wave functions of quantum many-body systems~\cite{Wu:2023fgp}.
Among many possible candidates, we consider the one along the line of the proposal in Ref.~\cite{Han:2019wue}, leaving other possibilities for future studies.
As the simplest large-$N$ gauge theory, we consider the SU($N$) two-matrix model, which is obtained by the dimensional reduction of $(1+2)$-dimensional SU($N$) Yang-Mills theory to $(1+0)$ dimensions, and its mass deformation.
Let us denote an NNQS by $\ket{\psi_\theta}$, where $\theta$ represents the parameters of the neural network.
By tuning the parameters appropriately, the energy $E_\theta\equiv\bra{\psi_\theta}\hat{H}\ket{\psi_\theta}$ can be minimized (here $\hat{H}$ is the Hamiltonian operator of the quantum system).
Such a state which minimizes the energy is a candidate for the ground state, in a variational sense.
Additionally, one has to check if the ground state can be efficiently expressed by a particular neural network under consideration, and if the correct minimum rather than a local minimum is obtained.
Of particular importance is if technical issues arise as the coupling and/or the matrix size $N$ increase, for example due to changing both $\hat{H}$ and the wave function of the ground state.
In previous works~\cite{Han:2019wue,Rinaldi:2021jbg,Han:2021our}, some signs of discrepancies between $E_\theta$ and the true ground state energy at strong coupling were observed.
To understand the situation better, we use lattice Monte Carlo simulations to determine the ground state energy at strong coupling so that we can cross-check the ground state energies obtained from the VMC method.
For the simpler case of $N=2$ it is possible to perform the exact diagonalization of a truncated Hamiltonian at large enough truncation level (such that the ground state energy is independent of it) and to obtain an accurate value for the ground state energy.
This is out of reach for matrix models with $N=3$ or larger, where only very low truncation levels could be numerically explored due to the exponentially large Hilbert space required, and that is why we resort to lattice Monte Carlo simulations based on stochastic sampling methods.

The Hamiltonian of the two-matrix model studied in this paper is
\begin{align}
\hat{H} = {\rm Tr}\left(
\frac{1}{2}\hat{P}_I^2
+
\frac{m^2}{2}\hat{X}_I^2
-
\frac{g^2}{4}[\hat{X}_I,\hat{X}_J]^2
\right) \ ,
\label{eq:bosonic_Hamiltonian}
\end{align}
where $I$ and $J$ runs from 1 to 2, and
\begin{align}
\hat{P}_I=\sum_{a=1}^{N^2-1}\hat{P}_I^a\tau_a \ ,
\qquad
\hat{X}_I=\sum_{a=1}^{N^2-1}\hat{X}_I^a\tau_a.
\label{bosonic_generator}
\end{align}
Here, $\tau_a$ are the generators of SU($N$) normalized as ${\rm Tr}(\tau_a\tau_b)=\delta_{ab}$.
The canonical commutation relation is
\begin{align}
[\hat{X}_{Ia},\hat{P}_{Jb}]
=
i\delta_{IJ}\delta_{ab} \ .
\label{canonical_commutation_relation_XP}
\end{align}
The Hamiltonian and the canonical commutation relation are invariant under the SU($N$) transformation $\hat{X}_{I,ij}\to (\Omega\hat{X}_{I}\Omega^{-1})_{ij}$, $\hat{P}_{I,ij}\to (\Omega\hat{P}_{I}\Omega^{-1})_{ij}$.
Typically, the physical states are restricted to singlets under this SU($N$) transformation (gauge singlets).
We denote the Hilbert space spanned by gauge singlets by ${\cal H}_{\rm inv}$.
We can also consider a bigger, extended Hilbert space ${\cal H}_{\rm ext}$ that contains gauge non-singlets.
Corresponding to the gauge singlet constraint, states in the extended Hilbert space connected by the SU($N$) transformation should be identified; this is the so-called gauge redundancy. 
Because operators $\hat{X}_I$ and $\hat{P}_I$ are not SU($N$) invariant, they are defined on ${\cal H}_{\rm ext}$.

The study of this two-matrix model can be a good starting point for various generalizations:
\begin{itemize}
    \item 
    The D0-brane matrix model~\cite{Banks:1996vh,Itzhaki:1998dd,deWit:1988wri} is obtained by increasing the number of matrices and by adding fermions. 
    This model gives a nonperturbative formulation of type-IIA superstring theory and M-theory. 
    The nature of the ground state will be crucial for the understanding of holographic emergent geometry~\cite{Hanada:2021ipb,Gautam:2022akq};
    \footnote{For the description of emergent geometry, $\mathcal{H}_{\rm ext}$ is more convenient than $\mathcal{H}_{\rm inv}$.}
    \item
    By using unitary variables instead of Hermitian variables, the Eguchi-Kawai model~\cite{Eguchi:1982nm} is obtained. 
    In the large-$N$ limit, the Eguchi-Kawai model is equivalent to the infinite-volume lattice gauge theory through the large-$N$ volume reduction correspondence;
    \item
    It is also possible to add spatial volume instead of going to the large-$N$ limit and study finite-$N$ Yang-Mills theory in nonzero spatial dimensions. 
    Note that Yang-Mills theory and QCD can be embedded into matrix models via the orbifold lattice construction~\cite{Buser:2020cvn,Bergner:2024qjl,Kaplan:2002wv}, hence the same simulation techniques can be used. 
\end{itemize}

This paper is organized as follows.
In Sec.~\ref{sec:Theory}, we discuss the Block Neural Autoregressive Flow (BNAF) ansatz for our simulations. 
Sec.~\ref{sec:VMC} discusses the parameter choices and results of these simulations. 
In Sec.~\ref{sec:lattice}, we perform lattice Markov Chain Monte Carlo (MCMC) simulations to compare with these results. 
Comparing the values obtained from variational Monte Carlo and lattice MCMC, we conclude that the former gives the correct ground state energy when the width of the neural network is sufficiently large. 
Sec.~\ref{sec:conclusions} contains a brief conclusion. 
Some additional details are deferred to the appendix~\ref{sec:t_dist_rev}.

\section{Variational Monte Carlo with Neural Network Quantum State} \label{sec:Theory}
In this section, we explain the VMC method with NNQS for the ground state wave function.
The architecture we use is the same as the one used in Ref.~\cite{Rinaldi:2021jbg,Han:2021our}, which is a simplified version of the one used in Ref.~\cite{Han:2019wue}.
\subsection{Variational Monte Carlo}
Firstly, let us explain the VMC method without specifying the details of the NNQS $\ket{\psi_\theta}$.
The wave function of the two-matrix model under consideration is the function of $2(N^2-1)$ real numbers $X_I^\alpha$, where $I=1,2$ and $\alpha=1,2,\cdots,N^2-1$.
The NNQS is expressed as a function $\psi_\theta(X)=\bra{X}\ket{\psi_\theta}\in\mathbb{C}$ parametrized by $\theta$, representing the amplitudes for a given vector of variables $X$.
Therefore, $|\psi_\theta(X)|^2$ is the probability distribution of $X\in\mathbb{R}^{2(N^2-1)}$.
For a given NNQS, the energy $E_\theta$ is obtained by 
\begin{align}
    E_\theta 
    = 
    \bra{\psi_\theta}\hat{H}\ket{\psi_\theta}
    =
    \int dX\ |\psi_\theta(X)|^2\cdot\frac{\bra{X}\hat{H}\ket{\psi_\theta}}{\psi_\theta(X)}\, . 
\end{align}
The energy $E_\theta$ is obtained by taking the average of $\epsilon_\theta(X)\equiv\frac{\bra{X}\hat{H}\ket{\psi_\theta}}{\psi_\theta(X)}$ with the probability distribution $|\psi_\theta(X)|^2$.
Symbolically, 
\begin{align}
    E_\theta 
    =  \mathbf{E}_{X\sim|\psi_\theta(X)|^2}[\epsilon_\theta(X)]. 
  \label{energy-average}
\end{align}
We minimize $E_\theta$ to find a candidate for the ground state wave function. 
For that purpose, we use the gradient descent method. 
Namely, we calculate the gradient $\nabla_\theta E_\theta$ with the method explained shortly, and update $\theta$ as 
\begin{align}  \theta\to\theta'=\theta-\beta \nabla_\theta E_\theta,
\label{eq:theta_update}
\end{align}
where $\beta$ is the learning rate. 

The wave function obtained in this way is not necessarily SU($N$)-invariant. 
The ground state wave function is SU($N$)-invariant~\cite{Hanada:2020uvt}, and for $N=2$ and $N=3$ it was observed numerically that the SU($N$)-invariant wave function consistent with the ground state is obtained~\cite{Rinaldi:2021jbg} indeed up to moderately large couplings. 
However, as we will see later, at larger $N$, states not invariant under SU($N$) may be obtained during the variational procedure.
To avoid this, we add a soft constraint term $C\sum_a\hat{G}_a^2$ to the Hamiltonian:
\begin{align}
\hat{H}'=\hat{H}+C\sum_a\hat{G}_a^2.
\end{align}
Here, $\hat{G}_a=i\sum_{I,b,c}f_{abc}\hat{X}_{I}^b\hat{P}_I^c$ are the generators of the SU($N$) gauge transformation, where $f_{abc}$ is the structure constant of SU($N$) algebra. 
By choosing $C$ appropriately and minimizing $E'_\theta=\bra{\psi_\theta}\hat{H}'\ket{\psi_\theta}$, we can obtain the SU($N$)-invariant ground state.

This method is called variational `Monte Carlo' because of the way the energy and its derivatives are calculated.
The point is that the NNQS is chosen in such a way that $X$ can easily be generated with the probability $|\psi_\theta(X)|^2$, and hence the Monte Carlo integration with this probability weight can be conducted. 
Then, $E_\theta$ can be estimated by taking the average of $\epsilon_\theta(X)$ due to \eqref{energy-average}. 
The gradient $\nabla_\theta E_\theta$ can also be calculated via Monte Carlo, by using the following relation:
\begin{align}
\nabla_\theta E_\theta 
    &=
    \int dX\ (\nabla_\theta|\psi_\theta(X)|^2)\cdot\epsilon_\theta(X)
    +
    \int dX\ |\psi_\theta(X)|^2\cdot(\nabla_\theta\epsilon_\theta(X))
    \nonumber\\
    &=
    \int dX\ |\psi_\theta(X)|^2\cdot\frac{(\nabla_\theta|\psi_\theta(X)|^2)}{|\psi_\theta(X)|^2}\cdot\epsilon_\theta(X)
    +
    \int dX\ |\psi_\theta(X)|^2\cdot(\nabla_\theta\epsilon_\theta(X))
    \nonumber\\
    &=
\mathbf{E}_{X\sim|\psi_\theta(X)|^2}[(\nabla_\theta\log|\psi_\theta(X)|^2)\epsilon_\theta(X)]
    +    \mathbf{E}_{X\sim|\psi_\theta(X)|^2}[\nabla_\theta\epsilon_\theta(X)].
\end{align}
\subsubsection*{Reweighting method}
A convenient trick in numerical computation is \textit{reweighting} from another probability distribution, say $p_\eta(X)$. 
We can rewrite \eqref{energy-average} in a trivial manner, as 
\begin{align}
    E_\theta 
    =  \mathbf{E}_{X\sim p_\eta(X)}
    \left[
\frac{\epsilon_\theta(X)\cdot|\psi_\theta(X)|^2}{p_\eta(X)}
\right]\, . 
\end{align}
This way of calculating $E_\theta$ is called reweighting because the probability distribution $|\psi_\theta(X)|^2$ is obtained from another probability distribution $p_\eta(X)$ by multiplying a \textit{reweighting factor} $|\psi_\theta(X)|^2/p_\eta(X)$. 
The gradient $\nabla_\theta E_\theta$ can also be evaluated by reweighting. 

Although the reweighting can work for any probability distribution $p_\eta(X)$ in principle, in practice it is good to choose $p_\eta(X)$ in such a way that the reweighting factor is close to 1.\footnote{Otherwise we have so-called overlap problem. More precisely, to calculate $
\mathbf{E}_{X\sim |\psi_\theta(X)|^2}
    \left[F(X)\right]
=
\mathbf{E}_{X\sim p_\eta(X)}
    \left[
\frac{F(X)\cdot|\psi_\theta(X)|^2}{p_\eta(X)}\right]$ efficiently, it is desirable to take $p_\eta(X)$ such that $\frac{F(X)\cdot|\psi_\theta(X)|^2}{p_\eta(X)}$ does not fluctuate much when $X$ is sampled according to $p_\eta(X)$. 
}
For example, one can take $p_\eta(X)$ close to $|\psi_\theta(X)|^2$ by minimizing the Kullback–Leibler (KL) divergence between the two distributions
\begin{align}
    D_{\rm KL}(p_\eta||\ |\psi_\theta|^2)
    =
\mathbf{E}_{X\sim p_\eta(X)}
    \left[
\log\frac{p_\eta(X)}{|\psi_\theta(X)|^2}
\right]\, . 
\end{align}
Therefore, in the reweighting case, one can repeat the update of $\theta$ by \eqref{eq:theta_update} and that of $\eta$ via $\eta\to\eta-\beta'\nabla_\eta D_{\rm KL}(p_\eta||\ |\psi_\theta|^2)$ to minimize $E_\theta$ and $D_{\rm KL}(p_\eta||\ |\psi_\theta|^2)$ simultaneously. 
\subsection{Neural Network Architecture}
In this section, we will first explain the architecture of the neural networks we are using.
Let us split the wave function into an absolute value and a phase as
\begin{align}
    \psi(X)
    =  |\psi(X)|\cdot e^{i\varphi(X)}. 
\end{align}
We introduce neural networks for $|\psi(X)|$ and $\varphi(X)$ separately.
We note that $\varphi(X)$ is not important for this paper as the phase in the ground state can be assumed to be constant; therefore, we will neglect $\varphi(X)$ in the following for simplicity. 

Because $|\psi_\theta(X)|^2$ gives the probability distribution, we want to take it in a form convenient for the Monte Carlo integral in the VMC algorithm, i.e. a form that it is easy to sample from.
We use a neural network for this purpose, adopting the Block Neural Autoregressive Flow (BNAF) architecture~\cite{decao2019block}. 
We will only sketch the main ideas here and refer to Ref.~\cite{decao2019block} for details. 
Its fundamental idea is to prepare a normalizing flow with an easy to compute logarithm of the absolute determinant of the Jacobian.
A (neural) normalizing flow works by taking a simple distribution $p_0(\vec{z})$ for $\vec{z}\in\mathbb{R}^d$ with $d=2(N^2-1)$ in our case (say a multivariate Gaussian distribution) and defining a one-to-one map (bijective transformation)
\begin{align}
\vec{x}=\vec{f}_\theta(\vec{z})
\end{align}
by using a neural network, where $x_i=X_1^i$ and $x_{N^2-1+i}=X_2^i$ for $i=1,\cdots,N^2-1$. 
Then, because
\begin{align}
p_0(\vec{z})d\vec{z}
    =
    p_0(\vec{z})
    \left|
\det\partial_{\vec{z}}\vec{f}_\theta
    \right|^{-1}
    d\vec{x}\, , 
\end{align}
we can relate the map $\vec{f}_\theta$ and the wave function $\psi_\theta$ as 
\begin{align}
|\psi_\theta(\vec{x})|^2
    =
    p_0(\vec{z})
    \left|
\det\partial_{\vec{z}}\vec{f}_\theta
    \right|^{-1}\, . 
\end{align}
The derivative $\frac{\partial}{\partial\vec{x}}$ can also be written in terms of $\vec{z}$. 
Therefore, once a map $f_\theta:\vec{z}\to\vec{x}$ is given, we can calculate $E_\theta$ and $\nabla_\theta E_\theta$. 
In the same manner, we can introduce a map $g_\eta:\vec{z}\to\vec{x}$ to define $p_\eta(X)$ and use the reweighting method.    

\begin{figure}
    \centering
\includegraphics[trim = 0mm 0mm 0mm 0mm, clip, width=0.6\linewidth]{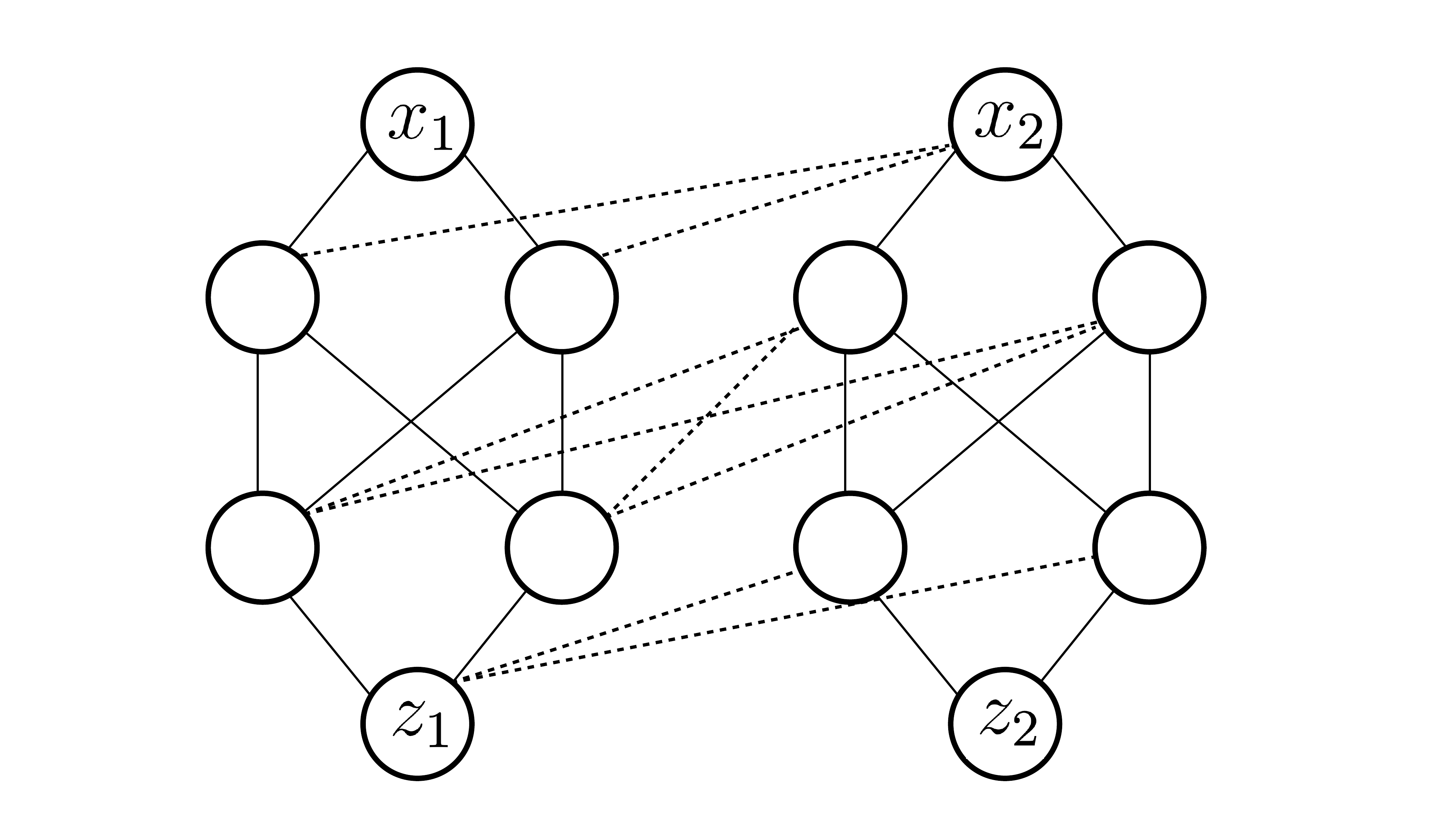}
    \caption{Architecture of a BNAF network depicted for $d=2$, $n=2$, $\alpha_1=\alpha_2=2$. The bottom layer corresponds to the input variables. Forwarding to the next layer is done via linear maps where links, both solid and dashed, indicate an allowed dependency. We observe that the output $x_1$ depends only on $z_1$, whereas $x_2$ depends both on $z_1$ and $z_2$. Dashed lines indicate a dependency across ``blocks'' (neurons inside a block are connected by solid lines). The network is allowed to have an arbitrary number of intermediate layers with an arbitrary number of neurons each. 
There is one block for each input-output pair $(z_i,x_i)$.}
    \label{fig:BNAF}
\end{figure}

To define $f_\theta$ and $g_\eta$, we use an autoregressive model from which it is easy to collect samples.
The starting point is to use the chain rule of conditional probability and express the joint probability as a product of conditional probabilities as
\begin{align}
    p(X) = p_{1} (x_1) \prod_{i=2}^{2(N^2-1)} p_{i | <i} (x_i | x_{<i})\, ,
\end{align}
where $x_{<i}$ refers to $\{x_1,\cdots,x_{i-1}\}$.
In an autoregressive model, firstly $x_1$ is obtained from $z_1$, then $x_2$ is obtained from $x_1$ and $z_2$ (equivalently, from $z_1$ and $z_2$), and then $x_3$ is obtained from $x_1$, $x_2$ and $z_3$ (equivalently, from $z_1$, $z_2$ and $z_3$), ..., and finally, $x_d$ is obtained from $x_1,\cdots,x_{d-1}$ and $z_d$. 
A schematic picture of the neural network we use in this work is shown in Fig.~\ref{fig:BNAF}. 
An important feature is that the weights connecting $z_l$ and $x_1,\cdots,x_{l-1}$ are masked. 
This autoregressive nature simplifies the computations because the Jacobian matrix $\partial_{\vec{z}}\vec{f}_\theta$ is lower-triangular. 
Specifically, the Jacobian is written as
\begin{align}
\det\partial_{\vec{z}}\vec{f}_\theta
=
\prod_{i=1}^d\frac{\partial f_i}{\partial z_i}\, . 
\end{align}
Therefore, the cost for computing the Jacobian is of order $d$, rather than a naive scaling $d^3$. 
Furthermore, the inverse of a matrix $\partial_{\vec{z}}\vec{f}_\theta$, which is needed to write $\partial_{\vec{x}}$ in terms of $\vec{z}$, can be calculated efficiently.  

The number of neurons in each layer is denoted by $\alpha_1,\cdots,\alpha_n$. 
The activation function is 
\begin{align}
  f_i(a_i)  \propto \text{Sinh} \left( (\text{Arcsinh}(a_i) + s_i) 
  \times\exp(t_i) \right)
\end{align}
where $i$ labels a neuron, $a_i$ is its input obtained via a linear map from the lower layer, and $s_i$ as well as $e^{t_i}$ are real parameters called skewness and tailweight.

As a default choice, we take $p_0(\vec{z})$ to be the Gaussian distribution, as in standard normalizing flows.
Later in this paper, we will see that the Student-t distribution can improve the performance of the architecture in some cases. 

The parameters $\theta$ that we vary to minimize the energy are the coefficients of the linear maps between the layers, as well as parameters of the activation functions that are applied after each intermediate layer.

\section{Ground state via variational Monte Carlo with Neural Network Quantum State}\label{sec:VMC}

\subsection{Physical parameter choices}
The Hamiltonian \eqref{eq:bosonic_Hamiltonian} has two parameters at fixed $N$: the coupling $\lambda = g^2 N$ and the mass $m$. 
Up to the rescaling of the variables, only the ratio $\lambda / m^2$ enters physical quantities. 
We consider three sets of parameters $(\lambda,m)$: 
\begin{itemize}
\item $\lambda = 0$, $m=1$:\\ Vanishing $\lambda$ at finite mass implies that the matrix degrees of freedom decouple and we are left with $2(N^2-1)$ uncoupled harmonic oscillators at unit mass. 
We use it as a sanity check because of its simple analytic solution. 
  
\item $\lambda = 1$, $m=0$:\\ Finite $\lambda$ at vanishing mass corresponds to the strong coupling regime of the model. 
This coincides with the bosonic part of the D0-brane matrix model where the number of matrices is increased from 2 to 9.
       
\item $\lambda = 1$, $m=1$:\\ Finite $\lambda$ at finite mass corresponds to the intermediate coupling regime. 
We study this parameter choice to check how the model interpolates between the two extreme cases above.
\end{itemize}

For applications in holography, one is usually interested in studying the large-$N$ extrapolation of the model. 
Therefore, we increase $N$ in the simulations enough to obtain a clear picture of the scaling of the results with $N$.
This will allow us to determine the usefulness of our method to study holography.
\subsection{Potential sources of errors}

\subsubsection{Hyperparameters}  \label{sec:Hyperparameters}
The hyperparameters determining our network structure are the width and the number of hidden layers (depth) in the BNAF architecture. 
We encode the network structure in a list $\alpha = [\alpha_1, \alpha_2, \ldots, \alpha_n]$ corresponding to $n$ hidden layers, where $\alpha_i$ are natural numbers. 
$\alpha_i$ corresponds to the number of hidden units in a single block at layer $i$ of the BNAF network structure \cite{decao2019block}, so there are $2(N^2-1)\alpha_i$ hidden units at layer $i$. 
A preliminary study we conducted in the early stages of this work showed that the results do not depend significantly on the number of hidden layers, at $\alpha_i\lesssim 10$. 
Therefore, we focused the case of only one hidden layer (i.e., take $n=1$) and studied wide range of $\alpha$.\footnote{
As we will see, to reproduce the ground-state energy precisely, we need to take $\alpha$ very large. With more than one layers, smaller width may be enough to achieve the same precision.  
}
\subsubsection{Control parameters}
We expect that the ground state is SU($N$)-invariant and rotational invariant. 
To confirm these, we monitored control parameters ($G\equiv\sum_a G_a^2$ and the angular momentum $R$) during simulations and confirmed they become close to zero, identifying a symmetric wave function.

Practically, to achieve the SU($N$) invariance with a good precision, we needed to choose a parameter $C$ for the gauge penalty term $C\sum_a\hat{G}_a^2$ appropriately. 
For $\alpha=2$, we studied several values of $N$ up to $N=11$. 
It turned out that for $N\leq 8$, $C=1$ works fine, while we need to increase $C$ to about $10$ above $N=8$. 
We also studied a wide range of $\alpha$ for $N=2,3,4$. 
There, $C=1$ was enough. 
The amount of breaking of the gauge-singlet constraint will be shown in Sec.~\ref{sec:vMC-results}, for $\alpha=2$ and $3\le N\le 11$. 
The rotational invariance followed without introducing a penalty term. 
\subsection{Simulation code}
We use the Python code already employed in Ref.~\cite{Rinaldi:2021jbg,Han:2021our} and added a minor extension to allow us to work with multiple intermediate layers. 
We refer to the GitHub repository \url{https://github.com/hanxzh94/minimal_BMN_MQM} for accessing to the code.
The code runs on a single Nvidia GPU for all the simulation parameters we employ.
\subsection{Simulation results}\label{sec:vMC-results}
In this section, we show the results of our VMC simulations. Firstly we perform a few sanity checks in Sec.~\ref{sec:Gaussian} and Sec.~\ref{sec:vMC-large-N}. 
In Sec.~\ref{sec:vMC-large-N}, we study a small fixed width (specifically, $\alpha=2$) and observe the disagreement with lattice simulations. 
As shown in  Sec.~\ref{sec:N2-large-alpha} and Sec.~\ref{sec:N3N4student}, this disagreement is resolved by taking the width sufficiently large. 
\subsubsection{Gaussian model: $\lambda =0, m=1$}\label{sec:Gaussian}
We start by comparing the case $\lambda=0$, $m=1$ to the analytic solution where the ground state energy $E = (N^2-1)$, i.e. the summed energy of $2(N^2-1)$ uncoupled harmonic oscillators in the ground state with the ground state energy $\frac{1}{2}$. Table \ref{Table:AnalyticSolution} shows the expected agreement for several values of $N$. 

\begin{table}[ht]
    \centering
    \begin{tabular}{|c||c|c|}
\hline
    $N$ & $H$: analytic solution & $H:$ measurement \\ 
    \hline
    \hline
    $3$ & $8$ & $8.01(3)$ \\ 
    $4$ & $15$ & $15.02(5)$ \\ 
    $6$ & $35$ & $35.1(3)$ \\
    $11$ & $120$ & $121.3(6)$ \\
    \hline
    \end{tabular}
\caption{$\lambda=0$, $m=1$, $\alpha=2$ measurements and analytic results for various $N$.
}\label{Table:AnalyticSolution}
\end{table}
The observed agreement is not surprising at all because $\vec{z}\to\vec{x}$ is simply the identity map. 
\subsubsection{Large $N$ at small $\alpha$}\label{sec:vMC-large-N}
\begin{figure}[htbp]
\centering
\includegraphics[trim = 0mm 0mm 0mm 14mm, clip, width=0.4\linewidth]
{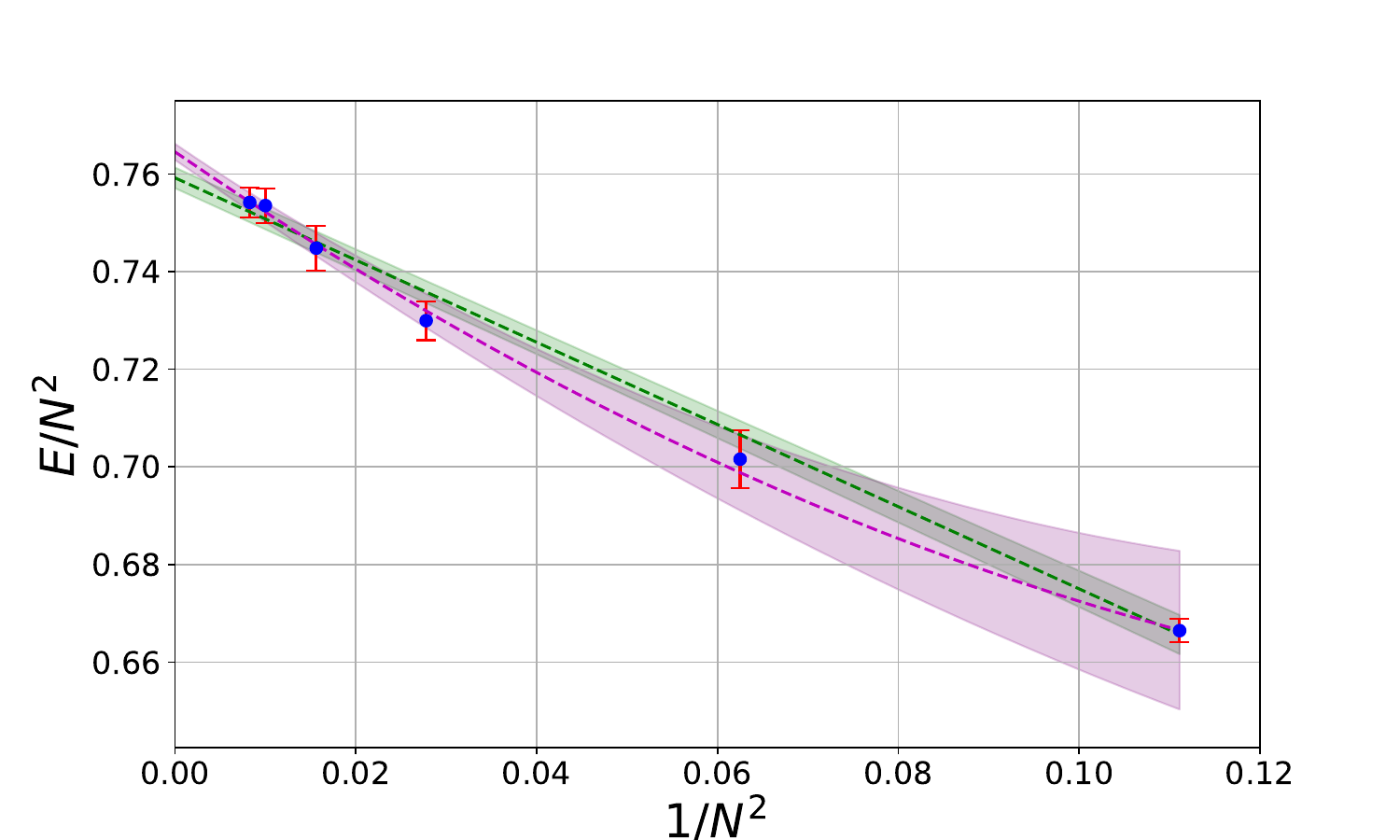}
\includegraphics[trim = 0mm 0mm 0mm 14mm, clip, width=0.4\linewidth]
{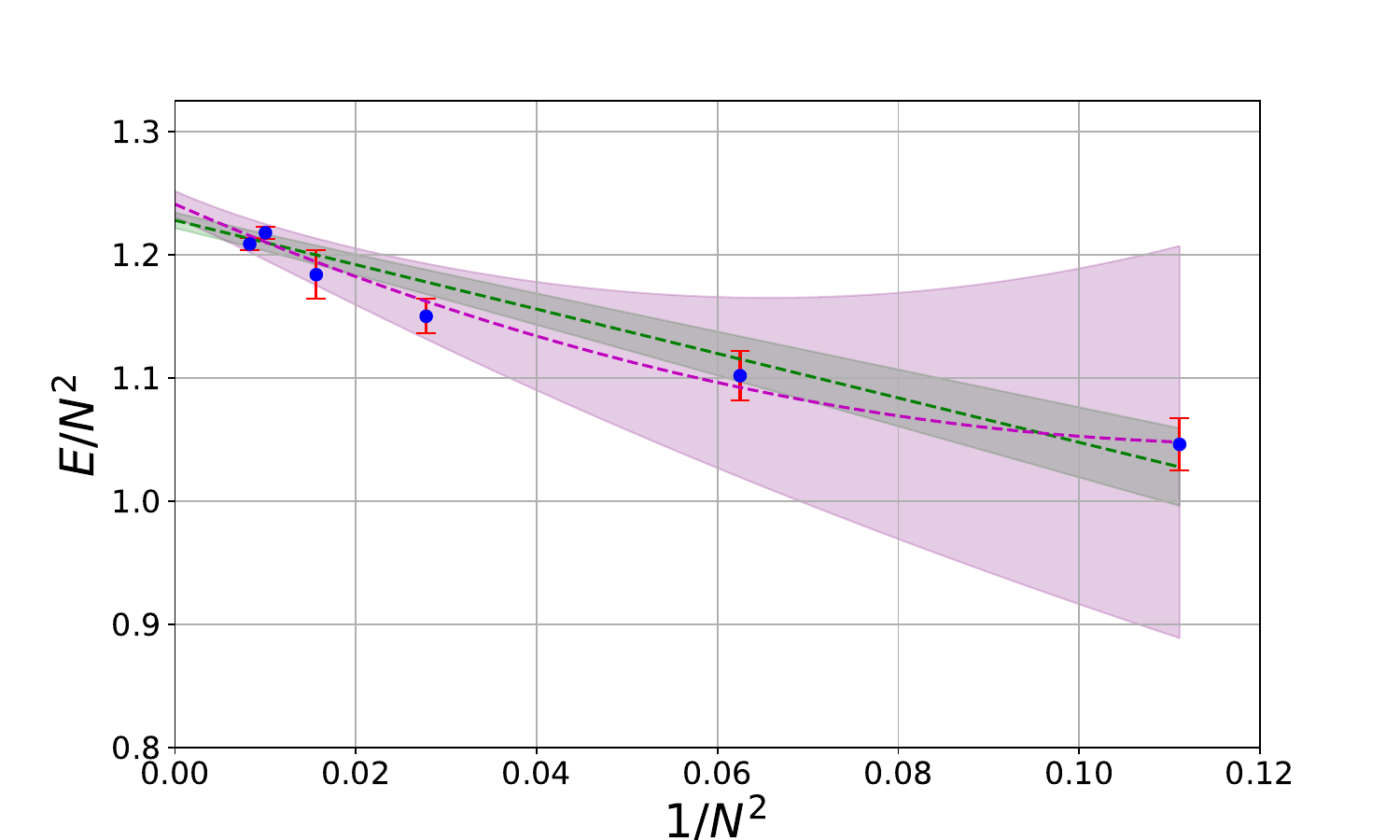}
\includegraphics[trim = 0mm 0mm 0mm 14mm, clip, width=0.4\linewidth]
{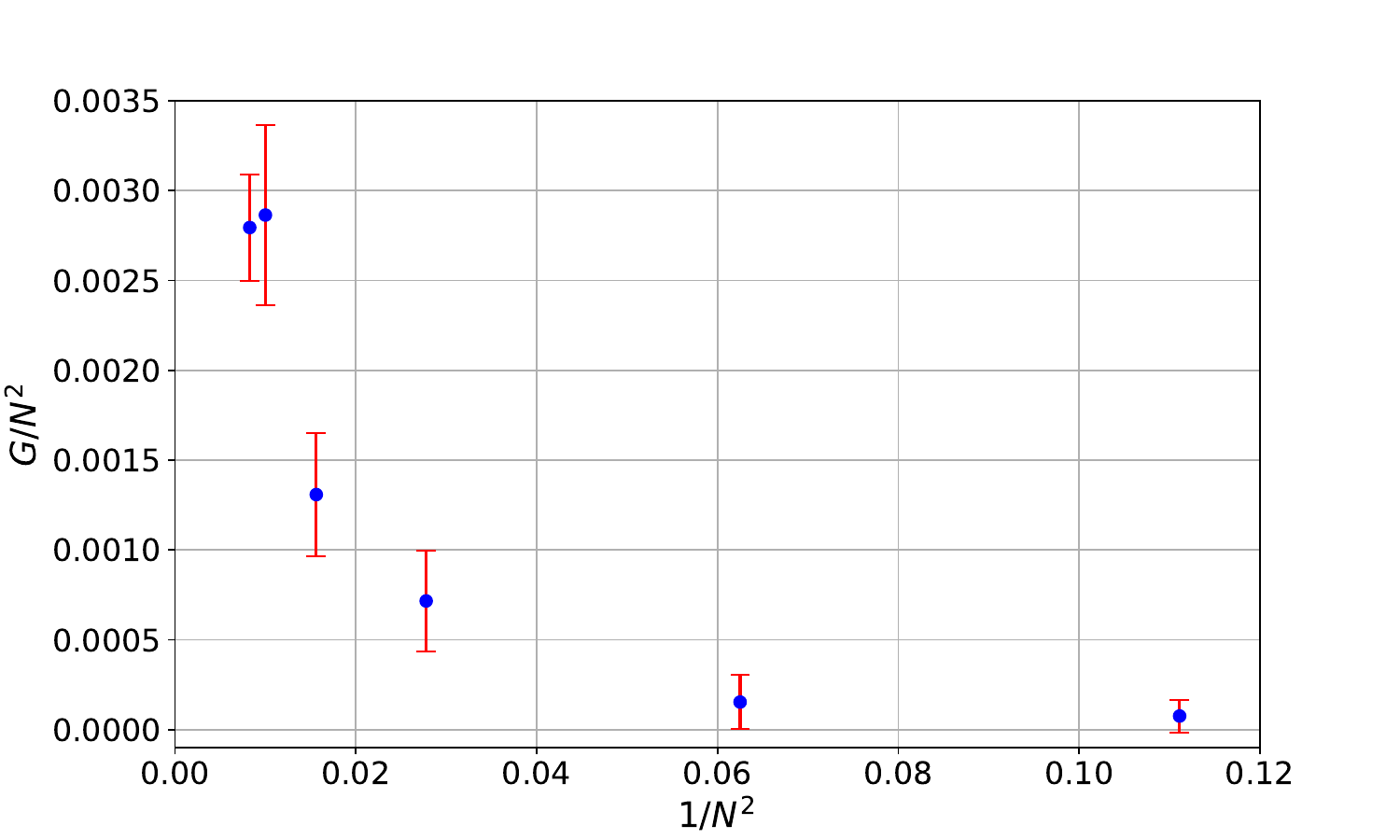}
\includegraphics[trim = 0mm 0mm 0mm 14mm, clip, width=0.4\linewidth]
{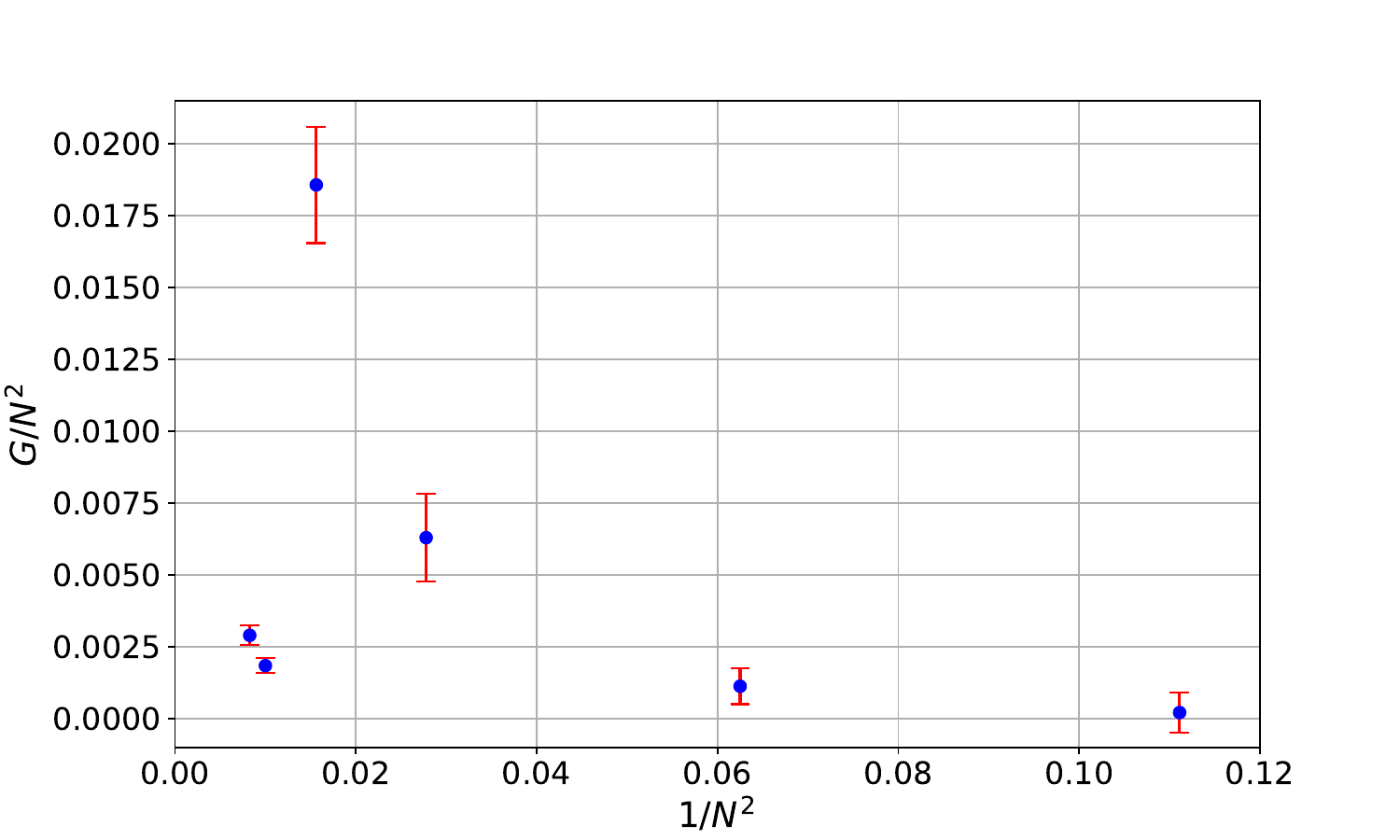}
\caption{ [Top] $E/N^2$ vs $1/N^2$ for $\lambda=1$, $m=0$ (top, left) and $\lambda=1$, $m=1$ (top, right). Linear and quadratic fit with respect to $1/N^2$ are shown. 
[Bottom ] $G/N^2$ vs $1/N^2$ for $\lambda=1$, $m=0$ (bottom, left) and $\lambda=1$, $m=1$ (bottom, right). 
The error bars for the data points are the standard deviation of the observables during the last 10 training epochs.
}\label{fig:large-N-small-alpha}
\end{figure}

Next, we studied the interacting case $\lambda =1$ both at $m=0$ and $m=1$ in the large $N$ limit at small $\alpha = 2$. 
The reasoning was to keep computational costs low (hence small $\alpha$) while trying to increase $N$ as much as possible. 
Specifically, we studied $N=3,4,6,8,10$, and $11$. 
We observed an $N$-dependence consistent with 't Hooft scaling. 
Specifically, by using a fit ansatz quadratic in $1/N^2$
\begin{align}
    \frac{E}{N^2}
    =
    \varepsilon_0
    +
\frac{\varepsilon_1}{N^2}
    +
\frac{\varepsilon_2}{N^4}\, 
\end{align}
we obtained 
\begin{align}
\varepsilon_0
&=
0.7645(16)
\nonumber\\
\varepsilon_1
&=
-1.26(10)
\nonumber\\
\varepsilon_2
&=
3.48(87)
\end{align}
for $\lambda=1$, $m=0$, 
and 
\begin{align}
\varepsilon_0
&=
1.241(10)
\nonumber\\
\varepsilon_1
&=
-3.2(10)
\nonumber\\
\varepsilon_2
&=
13.2(91)
\end{align}
for $\lambda=1$, $m=1$. 
See the top panels in Fig.~\ref{fig:large-N-small-alpha}, in which a linear fit is shown as well. 
As we can see the bottom panels in Fig.~\ref{fig:large-N-small-alpha}, the amount of the violation of gauge-singlet constraint characterized by $G/N^2$ is much smaller than 1 and well under control. 
We also observed that $R$ is consistent with zero, as expected for the ground state. 
However, the values of $E/N^2$ in the large-$N$ limit ($\varepsilon_0$ above) are larger than the Monte Carlo simulations discussed in Sec.~\ref{sec:lattice} (0.7039(11) and 1.1654(11) for $m=0$ and $m=1$, respectively) beyond the error bars. 
Below, we will argue that this disagreement is due to finite-$\alpha$ corrections.

\subsubsection{Large $\alpha$ at $N=2$}\label{sec:N2-large-alpha}
\begin{figure}[htbp]
\centering
\includegraphics[trim = 0mm 0mm 0mm 14mm, clip, width=0.7\linewidth]
{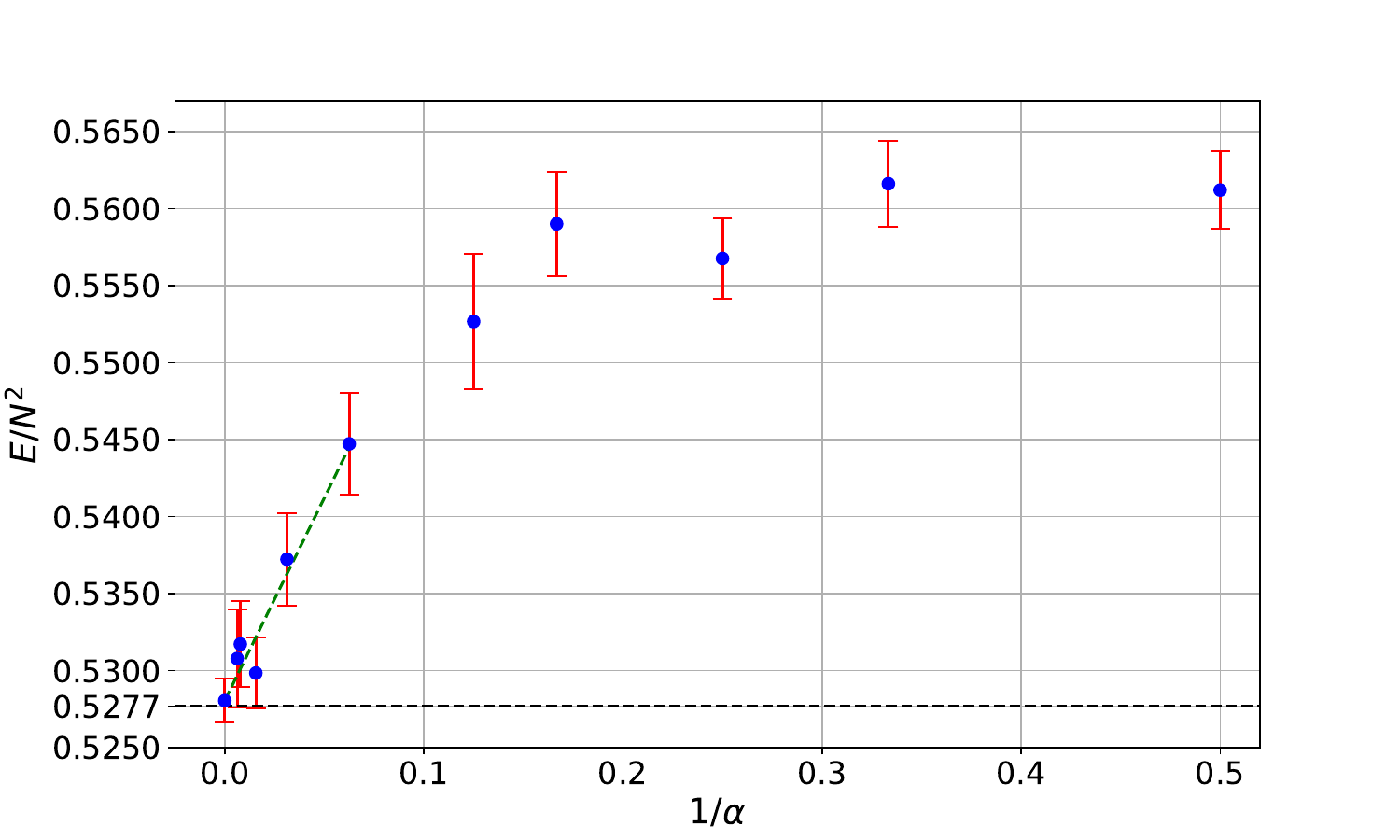}
\caption{Large $\alpha$ extrapolation of $E/N^2$ for $N=2$, $\lambda=1$, $m=0$ with single hidden layer. We observe that there is a dependence on $\alpha$ whose trend becomes approximately linear only for $\alpha > 10$. The point at $\alpha^{-1}=0$ is the linear extrapolation marked with the green line, which is 0.5280(14).
This is consistent with the correct value of the ground state energy obtained by exact diagonalization ($E/N^2=0.5277$), which is shown by the horizontal line.
The error bars for the data points are the standard deviation of the observables during the last 10 training epochs.
}\label{fig:LargeAlpha}
\end{figure}

To obtain an understanding of the influence of $\alpha$, we performed simulations for $\lambda=1, m=0$ at various $\alpha$ for $N=2$. 
Since $N=2$ is quite small, we were able to easily increase $\alpha$ to large values. 
Results are shown in Fig.~\ref{fig:LargeAlpha}. 
An important observation here is a sudden drop of the estimated ground state energy at $\alpha\gtrsim 10$. 
Large values of $\alpha$ are necessary to obtain the correct result and the approximately constant results at $\alpha < 10$ cannot be trusted to extrapolate to $\alpha = \infty$. 
We note that this limit is akin to a continuum limit in lattice gauge theory, as it allows us to approximate the wave function arbitrarily well. 

As a consequence, we are faced with a problem: quite large $\alpha$ may be required to obtain a sufficient approximation to the ground state energy, while simultaneously increasing $N$ and $\alpha$ to large values is expensive in terms of computational cost.

\begin{figure}[htbp]
\centering
\includegraphics[trim = 0mm 0mm 0mm 0mm, clip, width=1.0\linewidth]{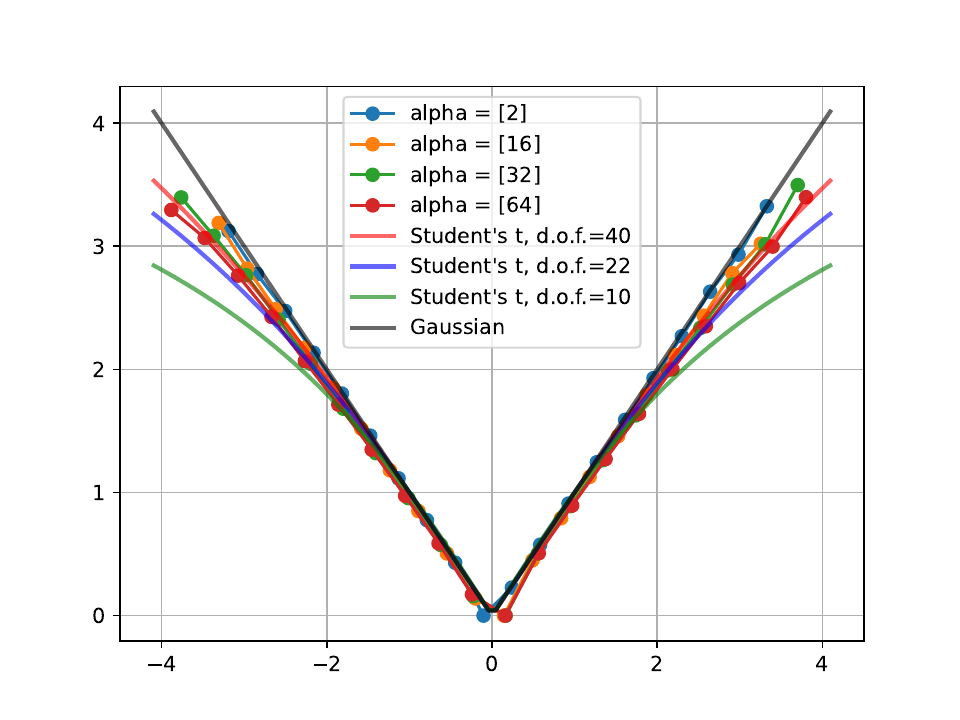}
\caption{We plot the wave function encoded by the neural network for $\lambda = 1$, $m=0$, $N=2$ for several $\alpha$ (line + dot). Furthermore, we plot a standard Gaussian and three examples of a Student's t distribution with various degrees of freedom (lines without dots). The plotted quantity is $\sqrt{- \log \left( \Psi(x_1) / \Psi(0))\right)}$, where we traced over all degrees of freedom but $x_1$. The purpose of this scaling is to visualize the deviation of the wave function in a single argument from a Gaussian, which looks like an absolute value (modulo constant scaling) when plotting the above quantity. \\
We observe that as $\alpha$ increases, the ``tail'' of the wave function becomes ``fatter'', i.e. it has larger values at the extremes. This effect increases as $\alpha$ increases, i.e. as the estimated ground state energy drops. We observe that a Student's t distribution with about $22$ degrees of freedom appears to be a good fit for the large $\alpha$ limit, certainly better than a Gaussian.}
\label{fig:wavefunctionfattail}
\end{figure}

To gain more insights into what is going on, we investigate the form of the wave function. 
For $N=2$, there are six variables. 
We define $\Psi(x_1)$ by integrating out five of them and kept only $x_1$, i.e., $\Psi(x_1)\equiv\int(\prod_{i=2}^6 dx_i)\psi(x_1,x_2,\cdots,x_6)$. 
In Fig.~\ref{fig:wavefunctionfattail}, $\sqrt{- \log \left( \Psi(x_1) / \Psi(0))\right)}$ is shown. 
If $\Psi(x_1)$ were Gaussian, then this would be proportional to $|x_1|$.
We observe that the wave function encoded in the large-$\alpha$ neural network has a fatter tail than a standard Gaussian and looks similar to a Student's t distribution at about $22$ degrees of freedom. 
Slow convergence to a fatter tail suggest that many fitting parameters are used to adjust the behavior at the tail part. 
As a consequence, it may be useful to change $p_0(\vec{z})$ from Gaussian to the Student's t distribution to achieve faster convergence to the correct ground state wave function. 
We study this issue in the next subsection. 

\subsubsection{$N=3$ and $N=4$: going to large $\alpha$ using Student's t Distribution}\label{sec:N3N4student}
In this subsection, we discuss the results of simulations conducted using the Student's t distribution as $p_0(\vec{z})$.
The t distribution is notably effective for datasets with significantly heavier tails than those typically modeled by the normal distribution.
This section explores the effectiveness of this choice on our BNAF architecture.
A concise review of the Student's t distribution is available in Appendix \ref*{sec:t_dist_rev} for further reference.
\begin{figure}[th!]
	\centering
	\includegraphics[width=\textwidth]{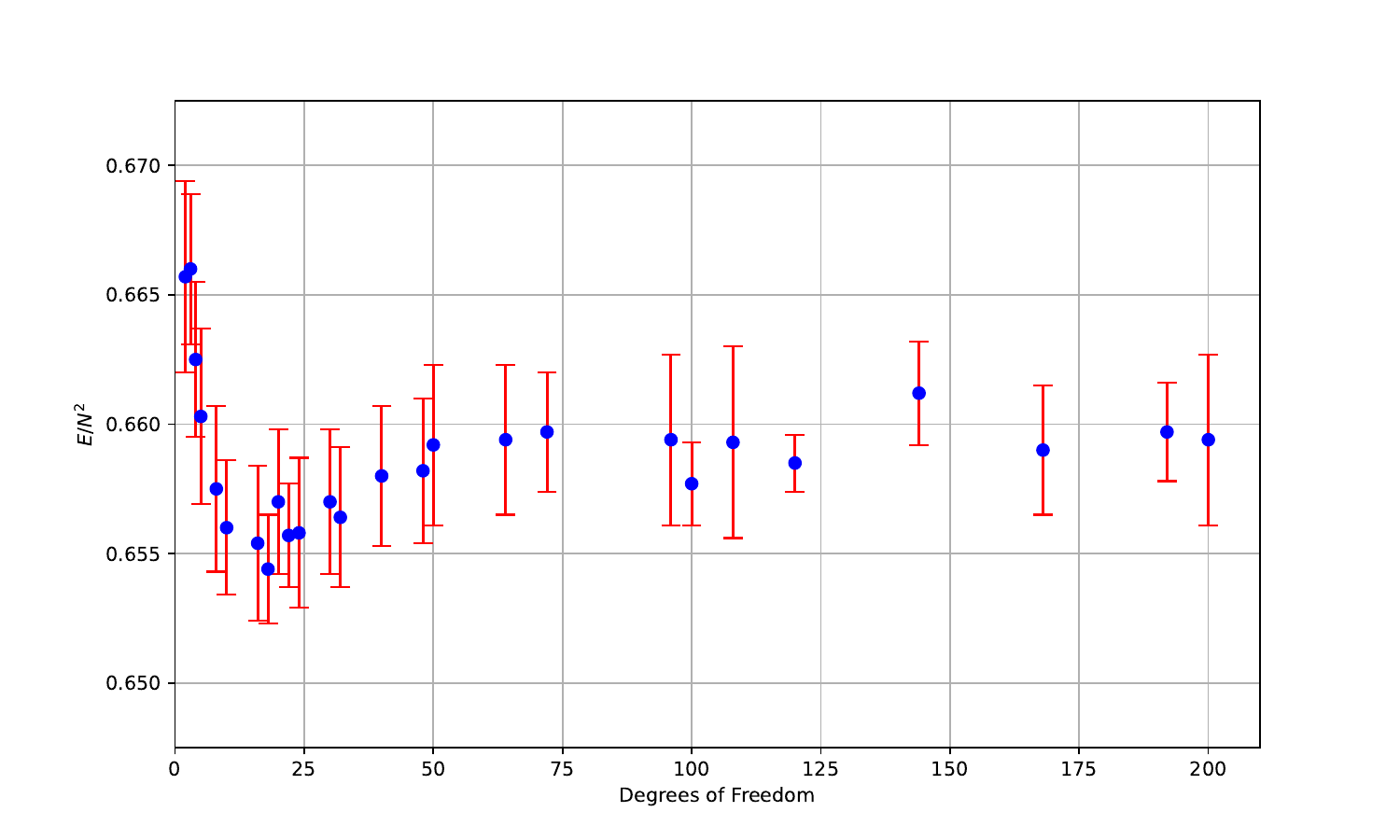}
	\caption{$E/N^2$ vs. degrees of freedom of t distribution for $N=3$, $\lambda=1$, $m=0$, $\alpha=15$.
 The error bars for the data points are the standard deviation of the observables during the last 10 training epochs.
 }
	\label{fig:E_v_dof_n3}
\end{figure}

\begin{figure}[ht!]
	\centering
	\includegraphics[width=\textwidth]{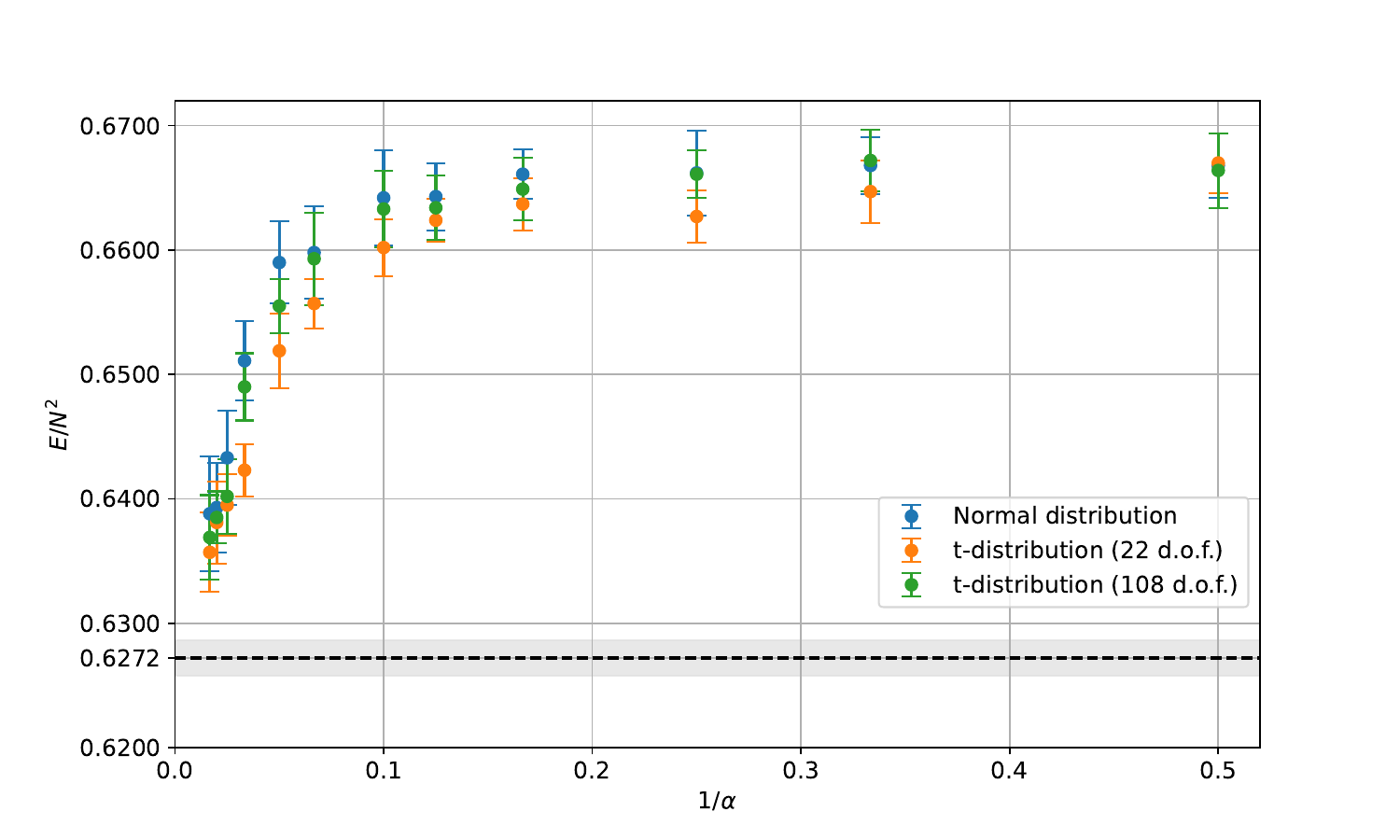}
	\caption{$E/N^2$ vs. $1/\alpha$ for different distribution ansatz at $N=3$, $\lambda=1$, $m=0$, one hidden layer.
At sufficiently large $\alpha$, the ground state energy is consistent with the one obtained through the lattice simulation, $E/N^2=0.6272(14)$, shown by the horizontal line and error band (see Sec.~\ref{Sec:lattice-N3N4}).
The error bars for the data points are the standard deviation of the observables during the last 10 training epochs.
 }
	\label{fig:LargeAlphaExtrCompr}
\end{figure}
\begin{table}[h]
	\centering
	\begin{tabular}{|c||c|c|c|}
        \hline
		$\alpha$ & t dist. (22 dof) & t dist. (108 dof) & Normal dist. \\
		\hline
\hline 
		2  & 0.6670(24) & 0.6664(30) & 0.6668(26) \\
		3  & 0.6647(25) & 0.6672(25) & 0.6668(23) \\
		4  & 0.6627(21) & 0.6661(19) & 0.6662(34) \\
		6  & 0.6637(21) & 0.6649(25) & 0.6661(20) \\
		8  & 0.6624(17) & 0.6634(26) & 0.6643(27) \\
		10 & 0.6602(23) & 0.6633(31) & 0.6642(38) \\
		15 & 0.6557(20) & 0.6593(37) & 0.6598(37) \\
		20 & 0.6519(30) & 0.6555(22) & 0.6590(33) \\
		30 & 0.6423(21) & 0.6490(27) & 0.6511(32) \\
		40 & 0.6395(25) & 0.6402(30) & 0.6433(38) \\
		50 & 0.6381(33) & 0.6385(21) & 0.6393(36) \\
		60 & 0.6357(32) & 0.6369(34) & 0.6388(46) \\
		\hline
	\end{tabular}
	\caption{$E/N^2$ at $N=3$, $\lambda=1$, $m=0$ is shown   
		for t distribution ansatz with $22$ and $108$ degrees of freedom, and normal distribution ansatz.
  True ground state energy obtained from lattice simulation is $0.6272(14)$.}
	\label{tab:comparison_ansatze}
\end{table}
\begin{figure}[htbp]
	\begin{subfigure}{0.5\textwidth}
	\scalebox{0.35}{\includegraphics{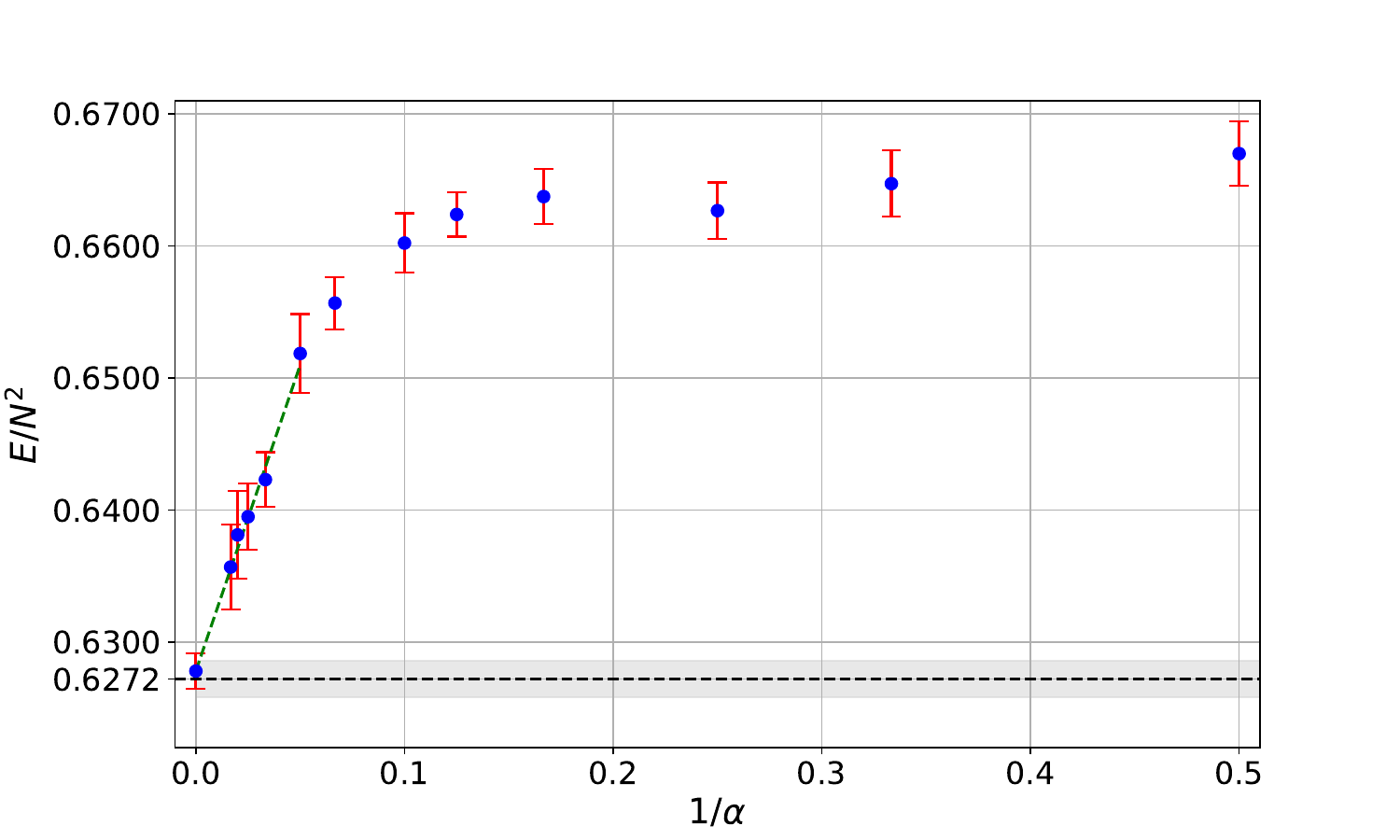}}
\caption{\mbox{$N=3$} }
\label{fig:HLargeAlpha_N3_t_dist}
	\end{subfigure}
	\begin{subfigure}{0.5\textwidth}
	\scalebox{0.35}{
\includegraphics{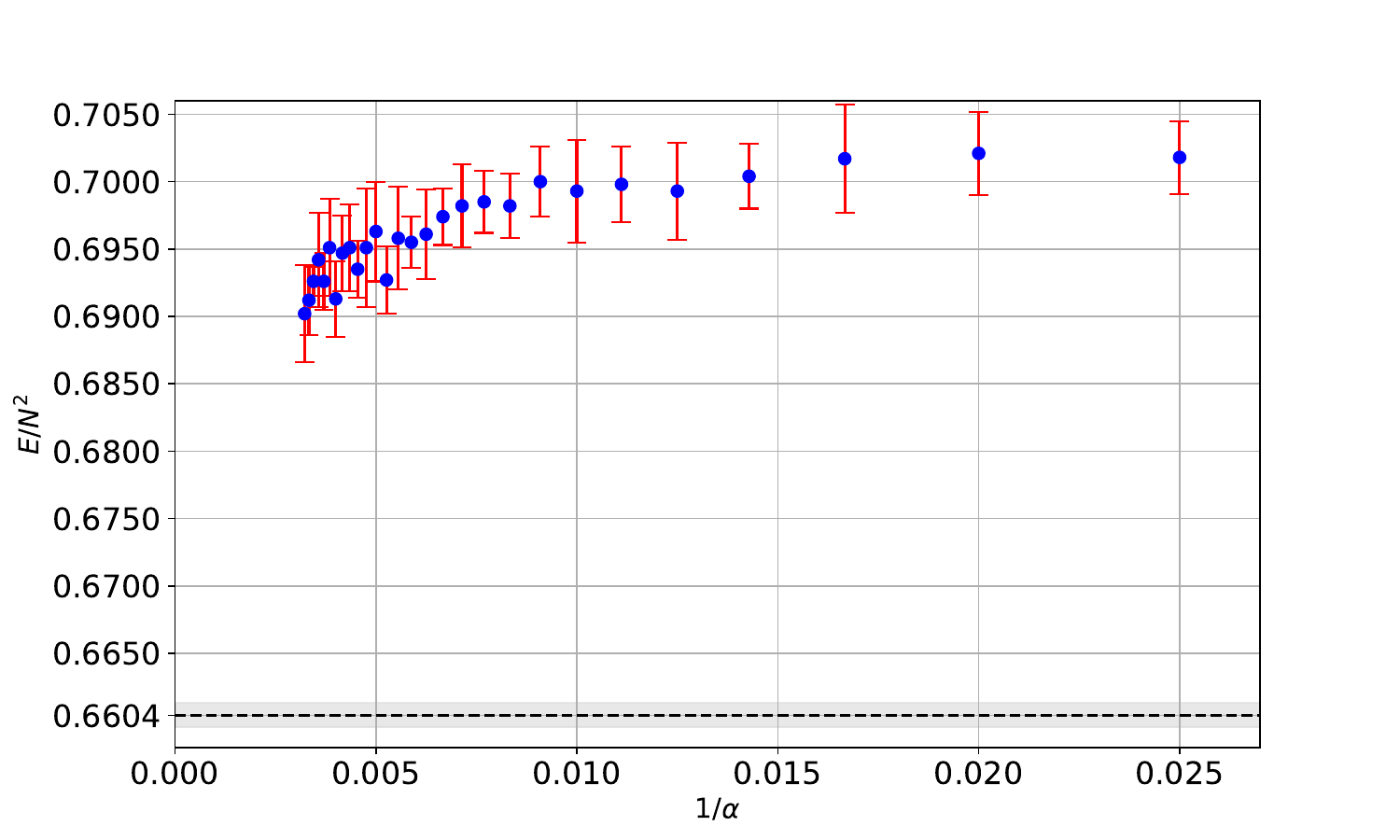}}
	\caption{\mbox{$N=4$}}
\label{fig:HLargeAlpha_N4_t_dist}
	\end{subfigure}
	\caption{ $E/N^2$ vs. $1/\alpha$ at $\lambda=1$, $m=0$, one hidden layer, for Student's t distribution ansatz with $22$ degrees of freedom.
 We also showed the value obtained through the lattice simulation by the horizontal line and error band. 
 For $N=3$, we showed the extrapolation $1/\alpha=0$. The extrapolated value $0.6278(13)$ is consistent with the lattice result, $0.6272(14)$.
For $N=4$, we can see a steady approach to the lattice result $E/N^2=0.6604(09)$ as $\alpha$ increases, although we need to study larger values of $\alpha$ to confirm a perfect agreement.  
 See Sec.~\ref{Sec:lattice-N3N4} for details of lattice simulations.
 The error bars for the data points are the standard deviation of the observables during the last 10 training epochs.
}\label{fig:HLargeAlpha_t_dist}
\end{figure}

Our simulation process began by examining various degrees of freedom within the t distribution to observe how they influence the variational energies. 
The results displayed in Fig.~\ref{fig:E_v_dof_n3} indicate that the normalized energy values are minimized around $18$ to $24$ degrees of freedom. 
Following an in-depth analysis where we examined the relationship between control parameters and degrees of freedom, we identified $22$ degrees of freedom as the most effective setting.

We conducted a set of simulations with parameters $N=3$, $\lambda=1$, $m=0$ using three distinct ansatz: the normal distribution, and t distributions with $22$ and $108$ degrees of freedom. 
The comparative results are visually represented in Fig.~\ref{fig:LargeAlphaExtrCompr}, where a decline in energy at higher alpha values is apparent, particularly with the t distribution at $22$ degrees of freedom.
These findings are quantitatively supported by the data in Tab.~\ref{tab:comparison_ansatze}, where the t distribution with $22$ degrees of freedom consistently outperforms the other ansatz at higher alpha levels.

Additionally, we extended our analysis to include simulations for $N=4$, keeping $\lambda=1$ and $m=0$ constant. 
The results, illustrated in Fig.~\ref{fig:HLargeAlpha_t_dist} for both $N=3$ and $N=4$, demonstrate a reduction in energy as $\alpha$ increases. 

We conclude that obtaining the large-$\alpha$ limit beyond $N=3$ is computationally more expensive.
Our results for $N=4$ are only partially satisfactory, and we did not attempt larger $N$.
Even though Student's t distribution helps to achieve convergence, rather large $\alpha$ is necessary at large $N$ to get a precise estimate of the ground state energy.  
\section{Lattice Monte Carlo for cross-check} \label{sec:lattice}
In this section, we describe the method used to obtain the lattice results, which are employed to cross-check the findings from the
variational Monte Carlo method discussed in Sec.~\ref{sec:vMC-results}.
The advantage of lattice Monte Carlo is that we are guaranteed to obtain the correct value of the ground state energy up to statistical and extrapolation errors that can be estimated systematically. 

The simulation code can be found at \url{https://github.com/masanorihanada/bosonic_matrix_model}.
\subsection{Simulation setup}
We use the Euclidean path integral to describe the finite-temperature theory.
The action is given by 
\begin{align}
S = N\int_0^\beta dt {\rm Tr}\left(
\frac{1}{2}(D_tX_I)^2
+
\frac{m^2}{2}X_I^2
-
\frac{\lambda}{4}[X_I,X_J]^2
\right) \ ,
\end{align}
where $\lambda=g^2N$.
The circumference of the temporal circle, $\beta$, is the inverse temperature $\beta=1/T$ of the system.
The matrices $X_I$ are traceless Hermitian.
$D_tX_I$ is the covariant derivative, $D_tX_I=\partial_tX_I-i[A_t,X_I]$, where $A_t$ is the gauge field.
This model has the confinement/deconfinement phase transition at critical temperature $T_{\rm c}$ which depends on $m^2$ and $\lambda$. It is convenient to use the 't Hooft expansion
\begin{align}
E(T)
=
\sum_{g=0}^\infty
N^{2-g}\varepsilon_g(T)
=
N^2\varepsilon_0(T)
+
\varepsilon_1(T)
+
\frac{\varepsilon_2(T)}{N^2}
+
\cdots, 
\label{eq:'tHooft-expansion}
\end{align}
where $\varepsilon_0$ is independent of temperature in the confined phase, i.e., $\varepsilon_0(T)$ is constant at $T<T_{\rm c}$. 

The lattice regularization used for this work~\cite{Rinaldi:2021jbg} is the tree-level improved action. 
Essentially the same regularization was used in Ref.~\cite{Berkowitz:2018qhn}.
The explicit form is
\begin{align}
S_{\rm lattice} = Na\sum_{t=1}^{n_t}{\rm Tr}
\left[
\frac{1}{2}(D_tX)_{I,t}^2
+
\frac{m^2}{2}X_{I,t}^2
-
\frac{\lambda}{4}[X_{I,t},X_{J,t}]^2
\right] \ ,
\end{align}
where $n_t$ is the number of lattice sites.
The lattice spacing $a$ and temperature $T=\beta^{-1}$ are related by $\beta=an_t$.
The covariant derivative $D_t$ is given by
\begin{align}
(D_tX)_{I,t} = \frac{1}{a}
\left[
-\frac{1}{2}U_t U_{t+a}X_{I,t+2a}U_{t+a}^\dagger U_{t}^\dagger
+
2U_t X_{I,t+a}U_{t}^\dagger
-
\frac{3}{2}X_{I,t}
\right] \ .
\end{align}
The periodic boundary condition $X_{I,n_t+1}=X_{I,1}$ is imposed.

In lattice Monte Carlo simulations, a sequence of lattice configurations
$\{X^{(1)},U^{(1)}\}\to \{X^{(2)},U^{(2)}\}
\to\cdots\to
\{X^{(k)},U^{(k)}\}\to\cdots$
is generated in such a way that their probability distribution converges to $\frac{1}{Z}e^{-S_{\rm lattice}[X,U]}$, where the normalization factor $Z$ is the partition function. 
For more details, see Ref.~\cite{Rinaldi:2021jbg}. 
At fixed $N$, it is necessary to take the continuum limit $a\to 0$ and zero-temperature limit $T=\frac{1}{an_t}\to 0$, which requires simulations with large lattice size $n_t$. 
If we focus on the large-$N$ limit, we can avoid this problem because $\varepsilon_0(T)$ is constant at $T<T_{\rm c}$ and hence we do not have to take the zero-temperature limit. 
 
\subsection{Simulation results}

\subsubsection{Large-$N$ limit}\label{sec:MCMC-large-N}
The large-$N$ limit of $\varepsilon_0\equiv\frac{E}{N^2}$ is independent of temperature in the confined phase.
This can be understood as a special version of the Eguchi-Kawai equivalence~\cite{Eguchi:1982nm}.
Therefore, we can use moderately high temperatures to estimate the ground state energy. 
Specifically, we use $T=0.8$ for $m^2=0, \lambda=1$ and $T=1$ for $m^2=1, \lambda=1$. 
This is a convenient property for numerical simulations because, if temperature is not low, we can control the discretization effect associated with a finite lattice spacing more easily. 

\begin{table}[ht]
    \centering
    \begin{tabular}{|c|c|c|c||c|c|c|c|}
    \hline
   \multicolumn{4}{|c||}{ $m^2=0, \lambda=1, T=0.8$}&
   \multicolumn{4}{|c|}{ $m^2=1, \lambda=1, T=1$}\\
    \hline
    \hline
    $N$ & $n_t$ & $E/N^2$ & $\langle |P|\rangle$ & $N$ & $n_t$ & $E/N^2$ & $\langle |P|\rangle$\\ 
    \hline
    8 & 16 & 0.6834(29) & 0.1998(23) & 8 & 16 & 1.1010(31) & 0.1649(17) \\
      & 20 & 0.6888(30) & 0.1959(21) &   & 20 & 1.1127(33) &  0.1608(17) \\
      & 24 & 0.6907(32) & 0.1965(22) &   & 24 & 1.1271(34) & 0.1625(16) \\
      & 32 & 0.7043(38) & 0.1933(21) &   & 32 & 1.1441(33) & 0.1646(17) \\
    \hline
    12 & 16 & 0.6662(15) & 0.1311(10) & 12 & 16 & 1.1017(21) & 0.1097(11) \\
     & 20 & 0.6751(16) & 0.1321(11) & & 20 & 1.1152(20) &
     0.1104(11) \\
     & 24 & 0.6823(15) & 0.1310(09) & & 24 & 1.1283(22) & 0.1083(11) \\
     & 32 & 0.6893(17) & 0.1284(11)& & 32 & 1.1361(23) & 0.1113(11) \\
     & 48 & 0.6978(22) & 0.1307(09) & & 48 & 1.1519(32) & 0.1084(12) \\
    \hline
    16 & 16 & 0.6598(09) & 0.09897(65) & 16 & 16 & 1.1022(17) & 0.08213(97) \\
     & 20 & 0.6679(10) & 0.09868(66) & & 20 & 1.1124(11) & 0.08163(61)\\
     & 24 & 0.6731(11) & 0.09644(79) & & 24 & 1.1227(19) & 0.08276(93)\\
     & 32 & 0.6831(14) & 0.09691(78) & & 32 & 1.1327(21) & 0.08337(92) \\
     & 48 & 0.6910(20) & 0.09678(84) & & 48 & 1.1471(16) & 0.08235(61)\\
    \hline
    24 & 16 & 0.6539(10) & 0.06575(72) & 24 & 16 & 1.0993(12) & 0.05482(65)\\
     & 20 & 0.6648(12) & 0.06591(75) & & 20 & 1.1125(08) & 0.05522(43)\\
     & 24 & 0.6716(12) & 0.06613(72) & & 24 & 1.1232(13) & 0.05470(62)\\
     & 32 & 0.6810(16) & 0.06448(71) & & 32 & 1.1344(16) & 0.05549(65)\\
     & 48 & 0.6913(17) & 0.06392(67) & & 48 & 1.1454(12) & 0.05411(48)\\
    \hline
    32 & 16 & 0.6532(06) & 0.04876(38) & 32 & 16 & 1.0997(09) & 0.04202(47) \\
     & 20 & 0.6632(07) & 0.04946(47) & & 20 & 1.1119(05) & 0.04049(31)\\
     & 24 & 0.6698(06) & 0.04834(43) & & 24 & 1.1221(10) & 0.04088(50)\\
     & 32 & 0.6784(08) & 0.04865(49) & & 32 & 1.1329(07) & 0.04137(33) \\
     & 48 & 0.6886(12) & 0.04772(51) & & 48 & 1.1438(11) & 0.04079(44) \\
    \hline
    \end{tabular}
    \caption{
   The summary of lattice simulation results at each $N$ and $n_t$. 
    }\label{Table:MCMC-summary}
\end{table}

\begin{figure}[h!]
\begin{subfigure}{.45\textwidth}
\rotatebox{0}{
\scalebox{0.45}{
\includegraphics{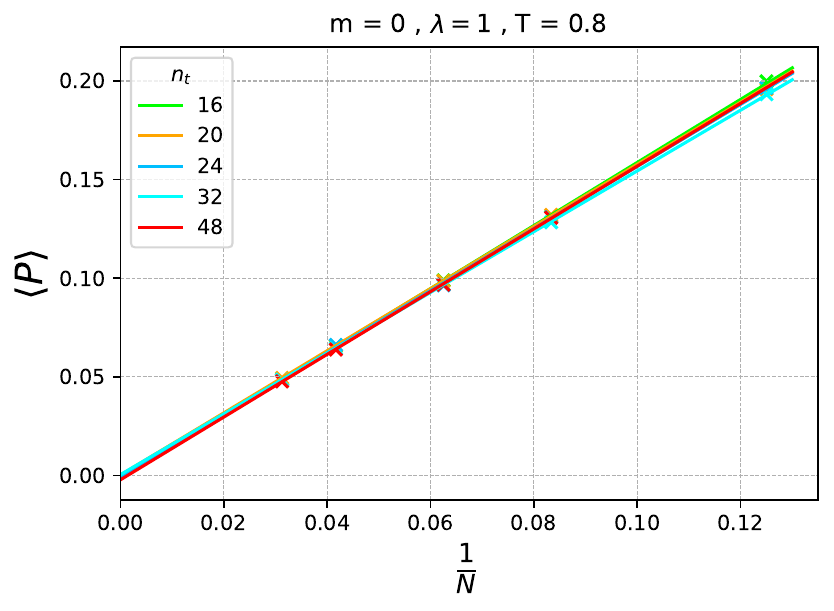}}}
\caption{\mbox{}}
\end{subfigure}
\begin{subfigure}{.45\textwidth}
\rotatebox{0}{
\scalebox{0.45}{
\includegraphics{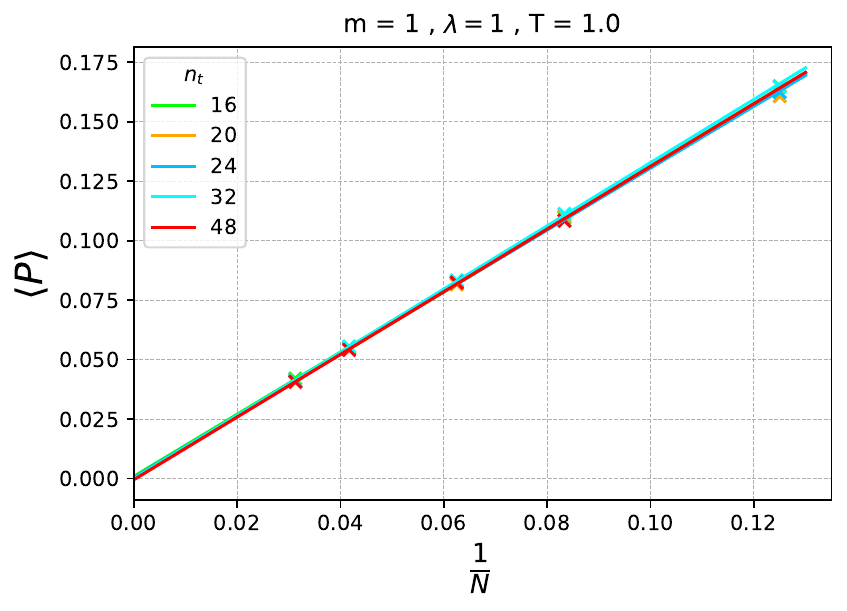}}}
\caption{\mbox{}}
\end{subfigure}
\caption{Large $N$ extrapolations of $\langle |P| \rangle$ at fixed $n_t$ for $\lambda=1$ and a) $m=0$, b) $m=1$, using fit parameters given in \ref{Table:Fixed nt Polyakov}. }
\label{fig:P vs 1/N}
\end{figure}

To confirm that we are indeed studying the confined phase, we use the Polyakov loop. 
In the confined phase, the Polyakov loop $P$ defined by $P=\frac{1}{N}{\rm Tr}(U_1U_2\cdots U_{n_t})$ should vanish. 
Specifically, at each fixed $n_t$, the expectation value of its absolute value $\langle |P|\rangle$ should scale as $\frac{1}{N}$ at sufficiently large $N$. 
In Fig.\ref{fig:P vs 1/N}, we plot $\langle |P|\rangle$ for several values of $n_t$ by taking the horizontal axis to be $1/N$. We can see that $\langle |P|\rangle$ vanishes proportionally to $1/N$ and that $\langle |P|\rangle$ is nearly independent of $n_t$. 
This shows that for each fixed number of lattice points, we are in the confined sector in the large $N$ limit. 
By using an ansatz $\langle |P|\rangle=p_0(n_t)+\frac{p_1(n_t)}{N}$ at each $n_t$, the values obtained for $p_0(n_t)$ and $p_1(n_t)$ are given in Tab.~\ref{Table:Fixed nt Polyakov} which are consistent with zero. 

\begin{table}[h!]
    \centering
    \begin{tabular}{|c|c|c||c|c|c|}
    \hline
   \multicolumn{3}{|c||}{ $m^2=0, \lambda=1, T=0.8$}&
   \multicolumn{3}{|c|}{ $m^2=1, \lambda=1, T=1$}\\
    \hline
    \hline
    $n_t$ & $p_0(n_t)$ & $p_1(n_t)$ & $n_t$ & $p_0(n_t)$ & $p_1(n_t)$\\ 
    \hline
     16 & -0.00112(72) & 1.598(15) & 16 & 0.00091(75) & 1.306(15) \\
     20 & 0.00030(80) & 1.575(15) & 20 & 0.000004(586) & 1.308(13)  \\
     24 & -0.00063(75) & 1.573(15) & 24 & 0.00045(76) & 1.300(14)  \\
     32 & 0.00050(82) & 1.539(16) & 32 & -0.00003(63) & 1.328(14)  \\
     48 & -0.00211(92) & 1.590(18) & 48 & -0.00046(88) & 1.317(19)   \\
    \hline
    \end{tabular}
    \caption{
   Linear extrapolations of Polyakov loop expectation values in $1/N$, at each fixed $n_t$, from HMC lattice simulations. 
    }\label{Table:Fixed nt Polyakov}
\end{table}

Next, let us estimate the large-$N$, continuum value of the energy $\varepsilon_0$. 
The 't Hooft expansion \eqref{eq:'tHooft-expansion} is applicable at each fixed $n_t$. 
Assuming that the coefficients admit the expansion with respect to $\frac{1}{n_t}$, we use the following ansatz
\begin{align}
\frac{1}{N^2}E(T,n_t)
=
\sum_{g=0}^\infty
\sum_{h=0}^\infty
\frac{\varepsilon_{g,h}}{N^{2g}n_t^h}
=
\varepsilon_{0,0}(T)
+
\frac{\varepsilon_{1,0}(T)}{N^2}
+
\frac{\varepsilon_{0,1}(T)}{n_t}
+
\frac{\varepsilon_{1,1}(T)}{N^2n_t}
+
\cdots.  
\label{eq:double-expansion}
\end{align}
We fit the above ansatz, truncated to linear order in $1/N^2$ and $1/n_t$, with 2D extrapolation using simulation results summarized in Tab.~\ref{Table:MCMC-summary}.  
The values we obtained are
\begin{align}
\varepsilon_{0,0}
&= 0.7039(11) 
\nonumber\\
\varepsilon_{1,0}
&= 1.06(31) 
\nonumber\\
\varepsilon_{0,1}
&= -0.848(23) 
\nonumber\\
\varepsilon_{1,1}
&= 15.9(65)
\label{energy-largeN-strong-coupling}
\end{align}
for $m^2=0, \lambda=1, T=0.8$, and
\begin{align}
\varepsilon_{0,0}
&= 1.1654(11)
\nonumber\\
\varepsilon_{1,0}
&= 1.26(33)
\nonumber\\
\varepsilon_{0,1}
&= -1.061(24) 
\nonumber\\
\varepsilon_{1,1}
&= -18.4(70) 
\end{align}
for $m^2=1, \lambda=1, T=1$. The values of $\varepsilon_{0,0}$ do not agree with the VMC results in Sec.~\ref{sec:vMC-large-N}. 

\begin{figure}[h!]
\begin{subfigure}{.45\textwidth}
\rotatebox{0}{
\scalebox{0.45}{
\includegraphics{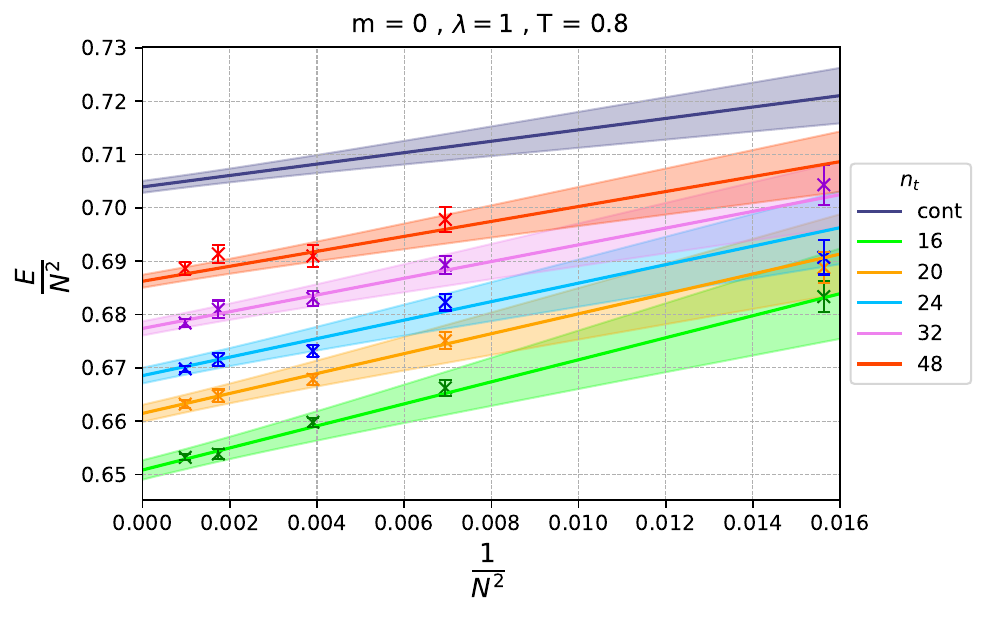}}}
\caption{\mbox{}}
\end{subfigure}
\begin{subfigure}{.45\textwidth}
\rotatebox{0}{
\scalebox{0.45}{
\includegraphics{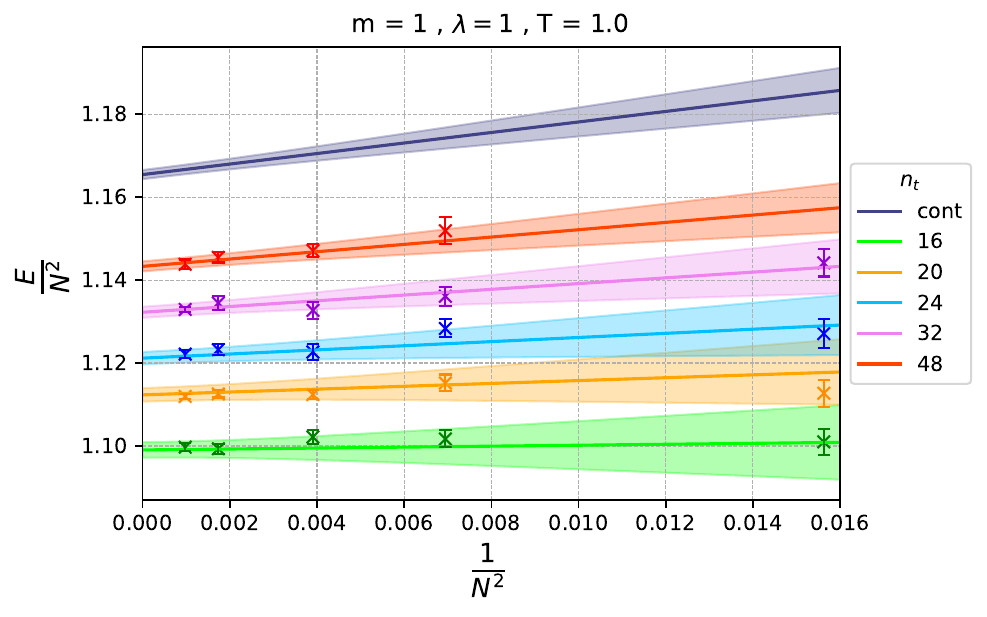}}}
\caption{\mbox{}}
\end{subfigure}
\caption{Large $N$ extrapolations of $E/N^2$ at fixed $n_t$ and in the continuum limit for $\lambda=1$ and a) $m=0$, b) $m=1$. }
\label{fig:E/N^2 vs 1/N^2}
\end{figure}

In Fig.~\ref{fig:E/N^2 vs 1/N^2}, we plot $E/N^2$ vs $1/N^2$ at fixed values of lattice sizes $n_t$, along with the continuum limit obtained using the ansatz in \eqref{eq:double-expansion}. 

\begin{figure}[h!]
\begin{subfigure}{.45\textwidth}
\rotatebox{0}{
\scalebox{0.45}{
\includegraphics{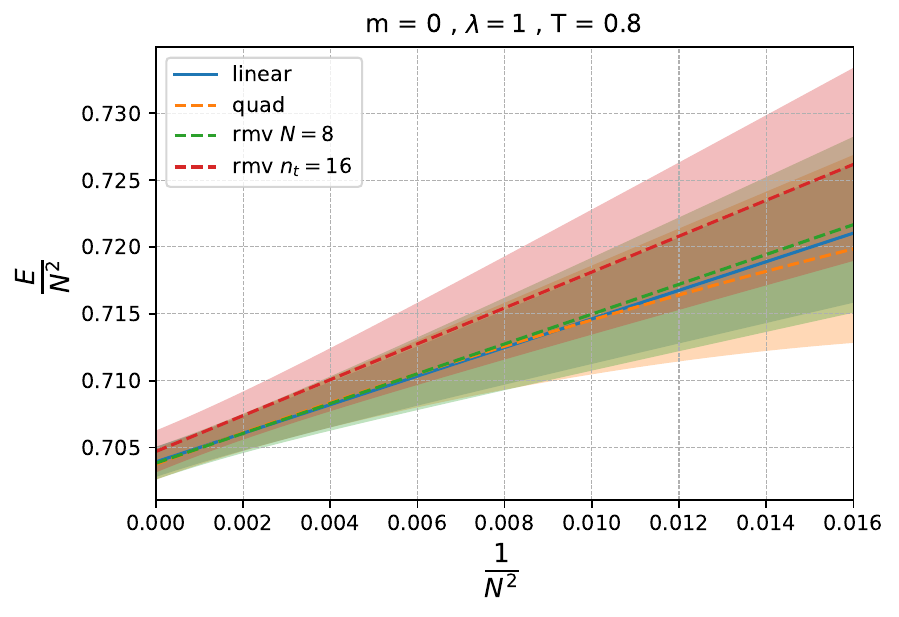}}}
\caption{\mbox{}}
\end{subfigure}
\begin{subfigure}{.45\textwidth}
\rotatebox{0}{
\scalebox{0.45}{
\includegraphics{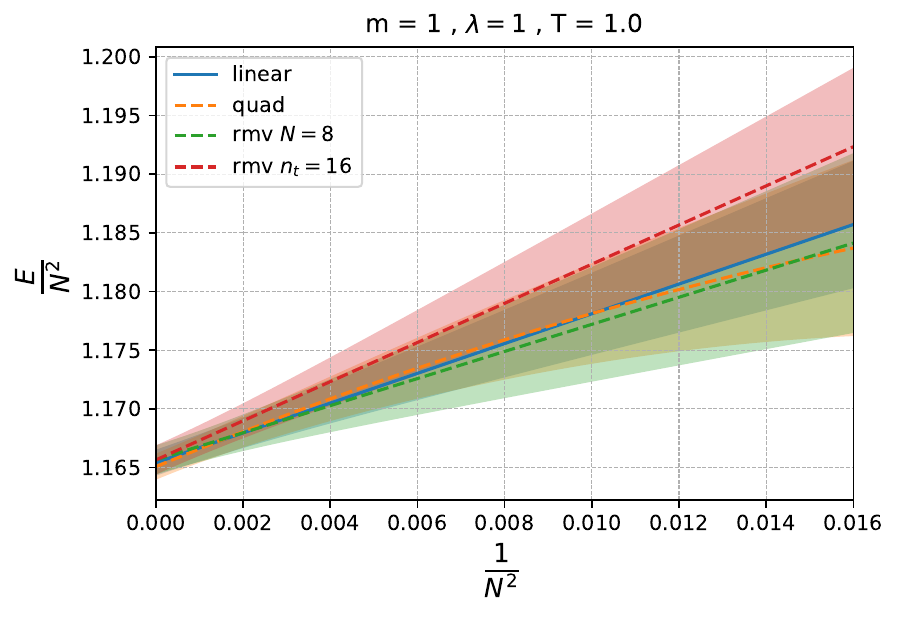}}}
\caption{\mbox{}}
\end{subfigure}
\caption{Large $N$ extrapolations of $E/N^2$ with linear ansatz, quadratic ansatz and removing certain values for $\lambda=1$ and a) $m=0$, b) $m=1$. }
\label{fig:stability analysis}
\end{figure}

In Fig.~\ref{fig:stability analysis}, we plot the continuum, large-$N$ extrapolation of energy comparing the plots obtained after considering all values of $N$ and $n_t$ with those obtained from $O(1/N^4)$ fit, removing $N=8$ or removing $n_t=16$. 
While the large-$N$ value is within error bars in all these cases, the continuum extrapolation obtained after removing $n_t=16$ values is not always contained within the error bars, indicating a small $n_t$ effect. 
To analyze this further, we study the small $N$ effect in Tab.~\ref{Table:Remove various N} by taking the large $N$ limit of energy after removing various values of $N$. 
To study the small $n_t$ effect, we remove the values at $n_t = 16$ and then take the large-$N$ limit of energy after removing various values of $N$ summarizing the results in Tab.~\ref{Table:Remove nt=16}. 
The difference is not big and hence does not resolve the mismatch with the VMC results.   

\begin{table}[h!]
    \centering
    \begin{tabular}{|c||c|c|c|c|c|c|}
    \hline
      & all $N$ & $N \neq 8$ & $N \neq 12$ & $N \neq 16$ & $N \neq 24$ & $N \neq 32$\\
      \hline
      \hline
     $m^2=0$ & 0.7039(11) & 0.7038(12) & 0.7043(11) & 0.7043(10) & 0.7029(12) & 0.7053(22) \\
     \hline
     $m^2=1$ & 1.1654(11) & 1.1657(12) & 1.1655(11) & 1.1655(11) & 1.1648(14) & 1.1661(19) \\
     \hline
    \end{tabular}
    \caption{
   Large $N$ values of $E/N^2$ at $\lambda=1$, after removing various values of $N$, with $n_t=16$. For $m^2 = 0$ the temperature is $T=0.8$ while for $m^2 = 1$ it is $T=1$. Terms up to linear order in $1/N^2$ and $1/n_t$ are included in the fits.
   }\label{Table:Remove various N}
\end{table}

\begin{table}[h!]
    \centering
    \begin{tabular}{|c||c|c|c|c|c|c|}
    \hline
      & all $N$ & $N \neq 8$ & $N \neq 12$ & $N \neq 16$ & $N \neq 24$ & $N \neq 32$\\
      \hline
      \hline
     $m^2=0$ & 0.7047(16) & 0.7050(17) & 0.7049(15) & 0.7048(14) & 0.7037(18) & 0.7060(30) \\
     \hline
     $m^2=1$ & 1.1656(12) & 1.1662(14) & 1.1655(01) & 1.1657(13) & 1.1653(15) & 1.1658(22) \\
     \hline
    \end{tabular}
    \caption{
   Large $N$ values of $E/N^2$ at $\lambda=1$, after removing various values of $N$, without $n_t=16$. For $m^2 = 0$ the temperature is $T=0.8$ while for $m^2 = 1$ it is $T=1$.
   Terms up to linear order in $1/N^2$ and $1/n_t$ are included in the fits.
}\label{Table:Remove nt=16}
\end{table}

\subsubsection{$N=3$ and $N=4$}\label{Sec:lattice-N3N4}
\begin{figure}[h!]
\begin{subfigure}{.45\textwidth}
\rotatebox{0}{
\scalebox{0.45}{
\includegraphics{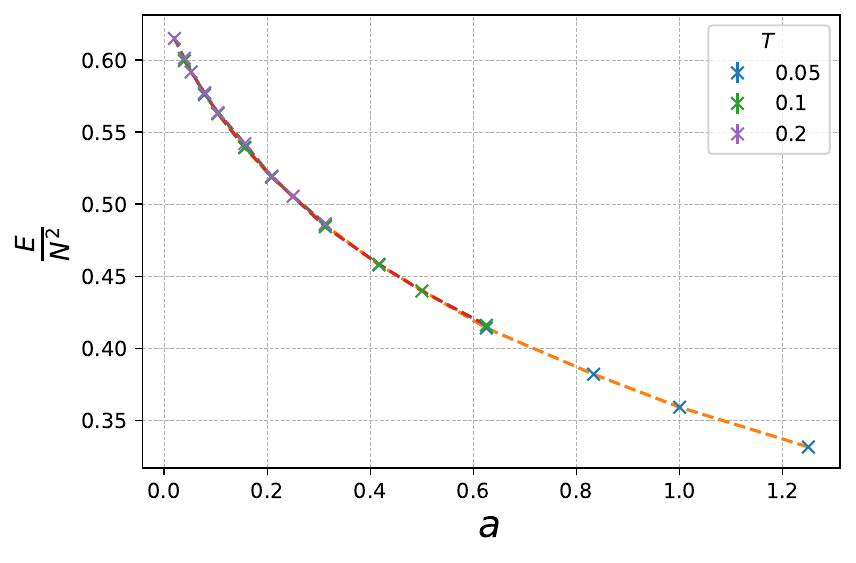}}}
\caption{\mbox{}}
\end{subfigure}
\begin{subfigure}{.45\textwidth}
\rotatebox{0}{
\scalebox{0.45}{
\includegraphics{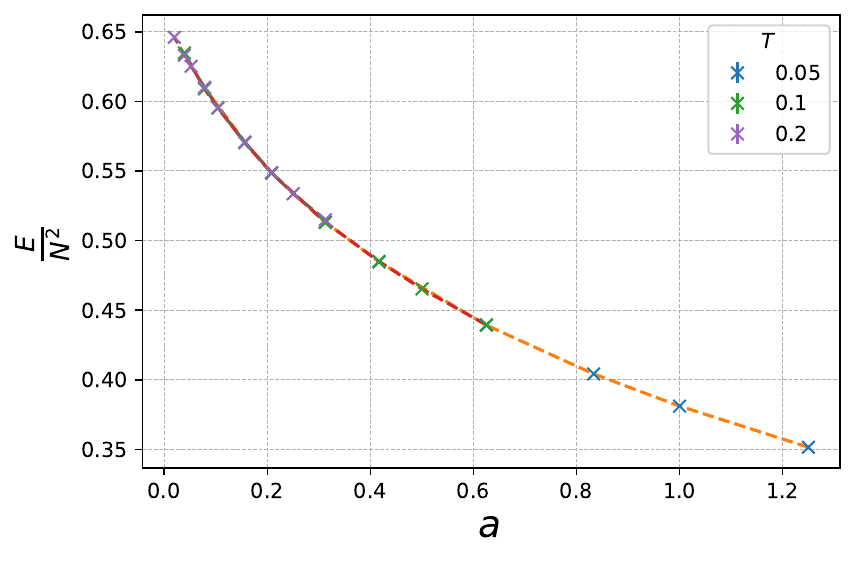}}}
\caption{\mbox{}}
\end{subfigure}
\caption{$E/N^2$ vs $a$ at $T=0.05, 0.1$, and $0.2$ for strong coupling i.e. $\lambda=1$ and $m^2=0$ for (a) $N=3$ and (b) $N=4$. 
}
\label{fig:E-vs-a}
\end{figure}

\begin{figure}[h!]
\begin{subfigure}{.45\textwidth}
\rotatebox{0}{
\scalebox{0.45}{
\includegraphics{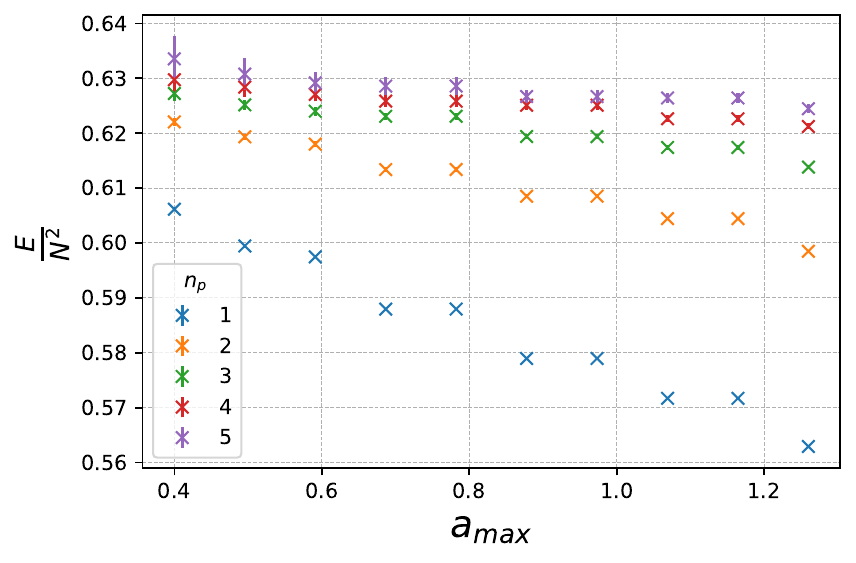}}}
\caption{\mbox{}}
\end{subfigure}
\begin{subfigure}{.45\textwidth}
\rotatebox{0}{
\scalebox{0.45}{
\includegraphics{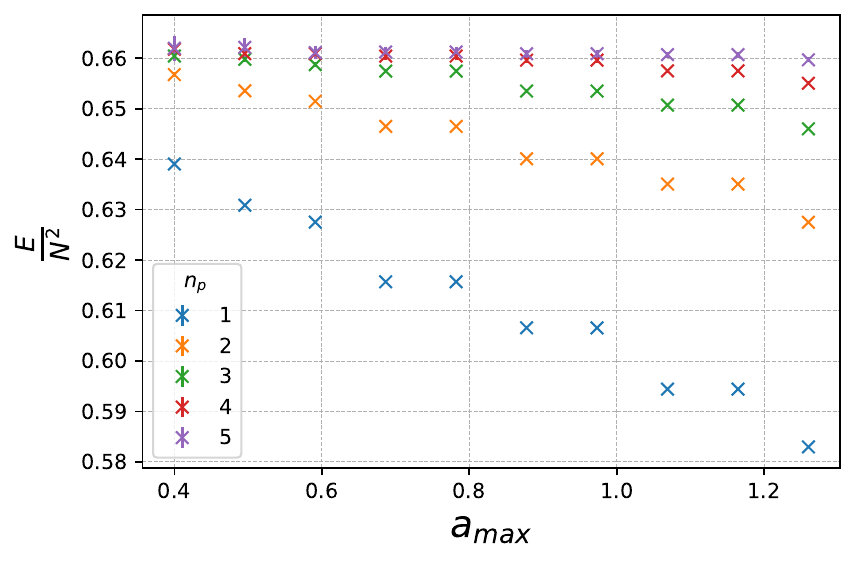}}}
\caption{\mbox{}}
\end{subfigure}
\caption{Results of systematic fitting using different data portions with polynomials of different order $n_p$ with $\lambda=1$ and $m^2=0$ for (a) $N=3$ and (b) $N=4$. $a_\text{max}$ on the horizontal axis is the cutoff value such that only data points with $a = \frac{1}{Tn_t} \leq a_{\text{max}}$ are considered for the energy fit.}
\label{fig:E_fit N34}
\end{figure}

We compute the energy for the bosonic two-matrix model for $N=3$ and $N=4$, to compare with the values obtained in Sec.~\ref{sec:N3N4student}. 
We report the results for strong coupling ($m^2=0$ and $\lambda = 1$) in the continuum limit.
We studied temperatures between $0.05 \leq T \leq 0.4$ and lattice sizes $n_t$ ranging from 16 to 256. 
The lattice spacing is given by $a = 1/(Tn_t)$. 
At each temperature $T$, the energy $E(T)$ is obtained by taking the continuum limit, $a\to 0$. 
In such a low-temperature region we consider, temperature dependence is expected to be exponentially small with respect to the inverse temperature, i.e., $E(T)\simeq E(T=0)+c\cdot e^{-c'/T}$, with order one constants $c$ and $c'$. 
In practice, we do not see $T$-dependence below a certain temperature. 
Indeed, as we can see from Fig.~\ref{fig:E-vs-a}, the energies at different temperatures line up as a function of lattice spacing $a$~\cite{Rinaldi:2021jbg} at $T \leq 0.2$.
Therefore, we use the data points at $T\leq0.2$ simultaneously to take the continuum limit, neglecting the dependence on $T$. 
We use the polynomial ansatz,
\begin{align}\label{Eq:Poly ansatz}
    E|_{a>0} = E|_{a=0} + \sum_{i=1}^{n_p} c_i a^i 
\end{align}
where $n_p$ is the degree of the polynomial and $c_i$ are fitting parameters. 
To study the systematic effects of the extrapolation to the continuum limit, we cut the data at different values of the maximum lattice spacing $a_\text{max}$, similarly to what was done in Ref.~\cite{Rinaldi:2021jbg}.
At each cut, we use only those data points that correspond to $a < a_{\text{max}}$. 
We also fit the functions using different degrees of polynomial $n_p$ with the ansatz in Eq.~(\ref{Eq:Poly ansatz}).
We repeat this fit for values of $a_\text{max} \in [0.4,1.25]$. 
The plot of the continuum extrapolation of the energy for different values of $a_\text{max}$ is shown in Fig.~\ref{fig:E_fit N34}. 
We summarize the values for some value of $a_\text{max}$ in Tab.~\ref{Table:E_fit values}. 
The fit is rather stable at $n_p\ge 3$ and near $a_{\rm max}=0.4$. Therefore, we take the values at  $n_p= 3$ and $a_{\rm max}=0.4$ as our lattice estimate:
\begin{align}
    \left.\frac{E}{N^2}\right|_{T=0} 
    =
    \left\{
    \begin{array}{cc}
         0.6272(14) &  (N=3)\\
         0.6604(09)& (N=4)
    \end{array}
    \right.
    \label{energy-N3N4-strong-coupling}
\end{align}

\begin{figure}
\centering
\scalebox{0.7}{
\includegraphics{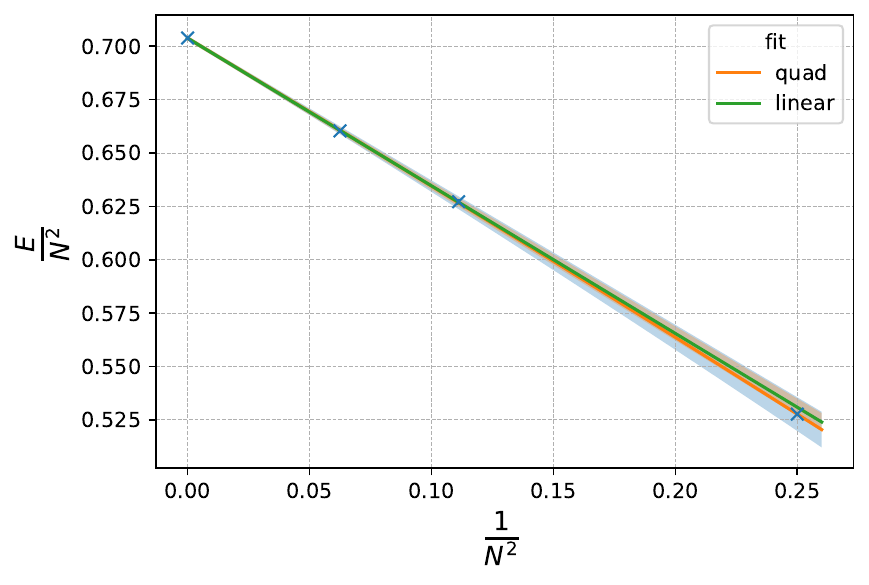}}
\caption{$E/N^2$ at $T=0$ for $N=2,3,4$, and $\infty$ vs $1/N^2$. (See \eqref{energy-largeN-strong-coupling} and \eqref{energy-N3N4-strong-coupling}.) The linear fit in $\frac{1}{N^2}$ is obtained by using only $N=3,4,\infty$ while the quadratic fit in $\frac{1}{N^2}$ uses all the points. We can see that $N=2$ value is not lined with the linear fit but is aligned with the others when we take O($1/N^4$) terms into account.}\label{fig:finite-N-correction}
\end{figure}

In Fig.~\ref{fig:finite-N-correction}, $E/N^2$ at $T=0$ for $N=2,3,4$, and $\infty$ are plotted against $1/N^2$. 
All data points line up on a straight line up to small corrections of order $1/N^4$, consistently with the 't Hooft scaling. 

\begin{table}[h!]
    \centering
    \begin{tabular}{|c||c|c||c|c|c|c|c|}
    
    \hline
    $N$ & $a_\text{max}$ & \# data points & $n_p = 1$ & $n_p = 2$ & $n_p = 3$ & $n_p = 4$ & $n_p = 5$ \\
      \hline
      \hline
     3 & 0.4 & 19 & 0.6062(03) & 0.6221(07) & 0.6272(14) & 0.6297(24) & 0.6335(42) \\
     & 0.4955 & 21 & 0.5995(03) & 0.6193(05) & 0.6252(10) & 0.6284(17) & 0.6307(29) \\
       & 0.5911 & 22& 0.5975(03) & 0.6180(05) & 0.6240(08) & 0.6271(13) & 0.6291(20)\\
       & 0.7822 & 24 & 0.5880(02) & 0.6134(04) & 0.6231(07) & 0.6258(11) & 0.6286(17)\\
       & 0.8778 & 25 & 0.5790(02) & 0.6085(03) & 0.6194(05) & 0.6251(08) & 0.6267(13) \\
       & 1.0689 & 26 & 0.5717(02) & 0.6044(03) & 0.6174(04) & 0.6226(06) & 0.6264(10) \\
       & 1.26 & 27 & 0.5629(02) & 0.5985(03) & 0.6138(04) & 0.6212(06) & 0.6245(08) \\
     \hline
     \hline
     4 & 0.4 & 19 & 0.6390(02) & 0.6568(05) & 0.6604(09) & 0.6618(15) & 0.6620(25) \\
       & 0.4955 & 21 &  0.6309(02) & 0.6536(04) & 0.6598(06) & 0.6609(11) & 0.6622(18) \\
       & 0.5911 & 22 & 0.6275(02) & 0.6515(03) & 0.6587(05) & 0.6609(08) & 0.6611(13)\\
       & 0.7822 & 24 & 0.6157(02) & 0.6465(03) & 0.6575(04) & 0.6604(07) & 0.6612(11)\\
       & 0.8778 & 25 & 0.6066(01) & 0.6401(02) & 0.6535(03) & 0.6596(05) & 0.6609(08) \\
       & 1.0689 & 26 & 0.5944(01) & 0.6351(02) & 0.6507(03) & 0.6575(04) & 0.6607(06) \\
       & 1.26 & 27 & 0.5830(01) & 0.6275(02) & 0.6460(02) & 0.6551(04) & 0.6597(05) \\
     \hline
    \end{tabular}
    \caption{
   Results of systematic fitting with using different data portions with polynomials of different order $n_p$ with $\lambda=1$ and $m^2=0$. 
    }\label{Table:E_fit values}
\end{table}

\section{Comparison between variational Monte Carlo and lattice Monte Carlo}
As for the large-$N$ limit, we recall that we obtained the continuum (effectively zero temperature) value $E/ N^2 = 0.7039(11)$ for the ground state energy in the large-$N$ limit for $m=0, \lambda=1$ as well as $E/ N^2 = 1.1654(11)$ for $m=1, \lambda=1$, from lattice Monte Carlo simulations. 
We use these values to check the performance of the variational Monte Carlo method based on NNQS. 

Let us focus on the more interesting case $m=0$ first, in which most simulation time was invested from the neural network perspective. 
As mentioned in Sec.~\ref{sec:vMC-large-N}, for a small width ($\alpha=2$) we observed that the scaling of the energy at large $N$ is consistent with the 't Hooft scaling and an extrapolation to $N=\infty$ provided us with $0.7645(16)$. 
This is clearly larger than the lattice value $0.7039(11)$ and inconsistent within errors, but so far we also neglected the corrections associated with the use of a small value of $\alpha$ (specifically, $\alpha=2$). 
For $N=2$ and $N=3$, we confirmed that the estimated energy goes down when larger values of $\alpha$ are used; see Fig.~\ref{fig:LargeAlpha} and Fig.~\ref{fig:LargeAlphaExtrCompr}. 
For larger $N$, we expect similar corrections that push down the estimated energy, although this was unfortunately so far out of computational reach. 

The case $m=1$ similarly appears to be consistent between the two approaches. 
A detailed large-$\alpha$ limit was not studied here, but the ground state energies are closer together and will likely match for a detailed computation at large $\alpha$. 
Since this case is closer to the Gaussian limit, we also expect it to be easier.

\section{\label{sec:conclusions}Conclusions and Future Directions}
In this paper, we applied the variational Monte Carlo method with neural network quantum state to the bosonic two-matrix model.
The architecture we use in this paper is the same as the one used in Ref.~\cite{Rinaldi:2021jbg}, which is a simplified version of the one used in Ref.~\cite{Han:2019wue}.
We estimated the ground state wave function and ground state energy via VMC and compared the ground state energy with that obtained from lattice Monte Carlo simulation to test the validity of the variational method.
For matrix size $N=2$ and 3, we confirmed the validity of the variational method by taking the width of the neural network $\alpha$ sufficiently large while fixing the depth. 
It was observed that, to obtain complete agreement, good control over the large-$\alpha$ limit is needed.
This may be challenging for $N > 3$. 

A possible way forward is to use a different probability distribution $p_0(\vec{z})$ that is closer to the actual wave function. 
Guessing such a probability distribution requires detailed knowledge about the ground state wave function that may be obtained using other approaches.
Another possibility is to use deeper network at larger $\alpha$; in this paper we fixed $\alpha$ when we studied deeper networks.
Last but not least, an approach that invest more computational resources is also possible, albeit perhaps less exciting.

Overall, we find the VMC approach very promising. 
We hope to report further development in the near future, adding more matrices, going to larger $N$, or adding fermions.
\section*{Acknowledgments}
We thank Xizhi Han, Jack Holden, and Lukas Seier for discussions.  
NB was supported by an International Junior Research Group grant of the Elite Network of Bavaria.
VG thanks STFC for the Doctoral Training Programme funding (ST/W507854-2021 Maths DTP).
MH thanks the STFC for the support through the consolidated grant ST/Z001072/1.
MH and ER thank the Royal Society International Exchanges award IEC/R3/213026.

\appendix

\section{Student's t Distribution}
\label{sec:t_dist_rev}
The Student's t distribution is a fundamental distribution in statistics, particularly useful for inference in scenarios with small sample sizes or unknown population variance. Its development is credited to William Sealy Gosset, who published under the pseudonym ``Student" \cite{student1908probable}.
	
The probability density function (pdf) for the generalized Student's t distribution, characterized by $\nu$ degrees of freedom, scale parameter $\sigma$, and location parameter $\mu$, is described as follows \cite{blattberg2010comparison}
\begin{equation}
	g(x|\nu, \mu, \sigma) = \frac{1}{\sqrt{\pi\nu}} \frac{\Gamma\left(\frac{\nu+1}{2}\right)}{\Gamma\left(\frac{\nu}{2}\right)\sigma} \left[1 + \frac{1}{\nu} \left(\frac{x - \mu}{\sigma}\right)^2 \right]^{-\frac{\nu+1}{2}},
\end{equation}
where $\nu > 0$, $\sigma > 0$, $x$ is a real number, and $\Gamma$ represents the Gamma function. The pdf is similar to the normal distribution but with heavier tails, indicating a higher likelihood of extreme values.
The parameter $\nu$, or degrees of freedom, influences the heaviness of these tails. The variance of the t distribution is given by $\frac{\nu \sigma^2}{\nu-2}$ for $\nu > 2$, and is not defined for $\nu \leq 2$~\cite{ahsanullah2014normal}.
	
At one degree of freedom, the Student's t distribution simplifies to the Cauchy distribution, known for its fat tails and indefinite moments. As $\nu$ approaches infinity, it converges to the normal distribution, with its heavy tails becoming less evident \cite{finner2008asymptotic}.
Despite some similarities to the normal distribution, the Student's t distribution differs significantly in kurtosis, with positive excess kurtosis indicating heavier tails than the normal distribution. This characteristic makes it more suitable for cases with unknown population variance~\cite{premaratne2000modeling}.

\bibliographystyle{unsrt}
\bibliography{references}

\end{document}